\documentclass[a4paper,twocolumn,journal]{IEEEtran}
\IEEEoverridecommandlockouts

\usepackage{amsfonts, amssymb, amsmath, cite, enumerate, amsthm, bm, caption, subfig, psfrag, cases, mathtools,booktabs,makecell}
\usepackage{ifthen}
\usepackage[usenames,dvipsnames]{color}

\ifCLASSINFOpdf
   \usepackage[pdftex]{graphicx}
   \DeclareGraphicsExtensions{.pdf,.jpeg,.png}
\else
   \usepackage[dvips]{graphicx}
   \DeclareGraphicsExtensions{.eps}
\fi

\DeclareMathOperator*{\diag}{diag}

\DeclareMathOperator*{\tr}{tr}

\newcommand{\ip}[2]{\left\langle#1,#2\right\rangle}
\newcommand{\norm}[1]{\left\lVert#1\right\rVert}
\renewcommand{\(}{\left(}
\renewcommand{\)}{\right)}
\renewcommand{\[}{\left[}
\renewcommand{\]}{\right]}

\newboolean{draft}
\newcommand{\isdraft}[2]{\ifthenelse{\boolean{draft}}{#1}{#2}}

\isdraft{\usepackage{setspace}}{}                              
\isdraft{\usepackage[footnotesize]{caption}}{}                 
\isdraft{\usepackage{paralist}}{}                              

\def \R {\mathbb{R}}

\def \< {\langle}
\def \> {\rangle}

\def \vd {\bm{d}}
\def \ve {\bm{e}}

\def \vx {\bm{x}}
\def \vu {\bm{u}}

\def \vy {\bm{y}}

\def \vX {\bm{X}}
\def \vY {\bm{Y}}
\def \vA {\bm{A}}
\def \vB {\bm{B}}
\def \vC {\bm{C}}
\def \vD {\bm{D}}
\def \vH {\bm{H}}
\def \vM {\bm{M}}
\def \vN {\bm{N}}
\def \vG {\bm{G}}
\def \vW {\bm{W}}
\def \vL {\bm{L}}
\def \vR {\bm{R}}
\def \vS {\bm{S}}
\def \vI {\bm{I}}
\def \vV {\bm{V}}
\def \vU {\bm{U}}

\def \vQ {\bm{Q}}
\def \vZ {\bm{Z}}

\def \vphi {\bm{\phi}}

\def \vSigma {\bm{\Sigma}}
\def \vGamma {\bm{\Gamma}}

\def \vPhi    {\bm{\Phi}}

\theoremstyle{plain}
\newtheorem{theorem}{Theorem}
\newtheorem{proposition}{Proposition}
\newtheorem{corollary}{Corollary}
\newtheorem{definition}{Definition}

\newtheorem{lemma}{Lemma}

\theoremstyle{remark}
\newtheorem{remark}{Remark}

\begin{document}
\title{Matrix Completion with Prior Information via Maximizing Correlation}

\author{Xu~Zhang, Wei~Cui, and Yulong~Liu
	\thanks{X.~Zhang and W.~Cui are with the School of Information and Electronics, Beijing Institute of Technology, Beijing 100081, China (e-mail: connorzx@bit.edu.cn; cuiwei@bit.edu.cn).}
	\thanks{Y.~Liu is with the School of Physics,
		Beijing Institute of Technology, Beijing 100081, China (e-mail: yulongliu@bit.edu.cn).}
}%

%



\maketitle

\pagestyle{empty}  
\thispagestyle{empty} 

\begin{abstract}
This paper studies the problem of completing a low-rank matrix from a few of its random entries with the aid of prior information. We suggest a strategy to incorporate prior information into the standard matrix completion procedure by maximizing the correlation between the original signal and the prior information. We also establish performance guarantees for the proposed method, which show that with suitable prior information, the proposed procedure can reduce the sample complexity of the standard matrix completion by a logarithmic factor. To illustrate the theory, we further analyze an important practical application where the prior subspace information is available. Both synthetic and real-world experiments are provided to verify the validity of the theory.
\end{abstract}
%

%
\IEEEpeerreviewmaketitle



\section{Introduction}

Recovering a structured signal from a relatively small number of linear measurements has received a great deal of attention during the past few decades \cite{chandrasekaran2012convex, vershynin2015estimation, tropp2015convex, thrampoulidis2015recovering, vaiter2015low}. Typical examples of structured signals include sparse vectors and low-rank matrices. This problem arises in an incredibly wide range of applications throughout signal processing and machine learning, such as medical imaging\cite{otazo2015low}, radar imaging\cite{potter2010sparsity}, communication systems\cite{garcia2017direct}, pattern recognition\cite{wright2010sparse}, collaborative filtering \cite{srebro2010collaborative} and so on. Since this problem is highly ill-posed in general, a popular recovery procedure is to seek the solution with the desired structure that is consistent with the observations, leading to
\begin{equation} \label{StandardSignalRecovery}
\min _{\vx}\|\vx\|_{\mathrm{sig}} ~~~~\text { s.t. }~~ \vy=\vA \vx,
\end{equation}
where $\vy$ stands for the measurements, $\vA$ denotes the measurement matrix, and $\norm{\cdot}_{\mathrm{sig}}$ is a suitable norm (or convex function) which promotes the structure of the signal.

However, in many practical applications of interest, we can also acquire other prior knowledge of the desired signal in addition to structural information. For instance, in compressed sensing, besides the sparsity constraint, we might have access to the support information \cite{vaswani2010modified}, the weights of the components of the desired signal \cite{candes2008enhancing,tanaka2010optimal,khajehnejad2011analyzing,scarlett2013compressed,mansour2017recoverya,needell2017weighted}, or a similar copy of the original signal \cite{chen2008prior,weizman2015compressed,mota2017compressed,zhang2017compressed}. While in matrix completion, certain subspace information of the desired low-rank matrix might be available to us \cite{srebro2010collaborative,foygel2011learning,xu2013speedup,chiang2015matrix,eftekhari2018weighted}. In these scenarios, a key question to ask is how to use side information to improve the recovery performance of structured signals.

Recently, the present authors suggested a unified framework to incorporate prior information into structured signals recovery via maximizing the correlation between the desired signal $\vx^\star$ and the prior information $\vphi$ in \cite{zhang2017compressed,zhang2018recovery}
\begin{equation} \label{OurSignalRecovery}
\min _{\vx}\|\vx\|_{\mathrm{sig}}-\lambda\langle\vx, \vphi\rangle   ~~~~\text { s.t. } \vy=\vA \vx,
\end{equation}
where $\vphi$ is some kind of prior information of the desired signal, $\langle\cdot, \cdot\rangle$ denotes the inner product, and $\lambda \ge 0$ is the tradeoff parameter. The motivation behind this approach is very natural since if $\vphi$ is similar to $\vx^\star$, then they may be highly correlated. We also theoretically demonstrate that, under sub-Gaussian measurements, this approach \eqref{OurSignalRecovery} can greatly outperform the standard structured signal recovery procedure \eqref{StandardSignalRecovery} when the prior information is reliable.

When specialized this framework to low-rank matrices, the recovery procedure \eqref{OurSignalRecovery} becomes
\begin{equation}\label{OurMatrixRecovery}
\min_{\vX \in \R^{n \times n}}  \norm{\vX}_{*} - \lambda \ip{\vPhi}{\vX} \quad \text{s.t.}~\vy=\mathcal{A}(\vX),
\end{equation}
where $\norm{\cdot}_{*}$ is the nuclear norm, $\ip{\vPhi}{\vX} = \tr(\vPhi^T\vX)$ denotes the matrix inner product, and $\mathcal{A}: \vX \to \sum_{j=1}^{m} \ip{\vA^j}{\vX} \ve_j$ is the measurement operator. Here, $\ve_1,\ldots,\ve_m$ denote the standard basis vectors in $\R^{m}$ and $\vA^1,\ldots,\vA^m \in \R^{n \times n}$ are measurement matrices. Although the theory developed for \eqref{OurMatrixRecovery} under sub-Gaussian measurements (i.e., $\{\vA^j\}$ are independent sub-Gaussian matrices) is very informative, it might be far from the type of observations we often encounter in practice. The following are some typical practical applications where $\{\vA^j\}$ are highly structured (i.e., $\vA^j$ has only a single nonzero value of $1$ corresponding to the row and column of the observed element) and some prior information is available:
\begin{itemize}
 \item \textbf{Semi-supervised clustering}\cite{yi2013semi,bair2013semi}. Semi-supervised clustering is an important machine learning problem, which is to find a good clustering such that similar items belong to the same cluster based on a relatively small amount of labeled data. One promising approach is to construct a similarity matrix with missing entries by using the labeled data and then complete the partial similarity matrix via matrix completion. In addition, the data attributes can be collected as side information, which represent the similarity among items.
 \item \textbf{Collaborative filtering} \cite{koren2010collaborative,rao2015collaborative,xu2016dynamic}. Collaborative filtering is another promising machine learning problem, which is to predict new ratings based on a limited number of ratings for different movies from different users. One popular scheme is to construct a partial rating matrix based on the known ratings and then complete the partial user-item rating matrix by using matrix completion. Moreover, user attributes and item attributes can serve as prior information. Here, user attributes denote the similarities among users while item attributes illustrate the similarities among items.
 \item \textbf{Dynamic sensor network localization} \cite{so2007theory,wang2008further,vaghefi2012cooperative}. Dynamic sensor network localization is a key technology in sensor wireless network, which helps the battery-powered system to improve the location accuracy and network efficiency. Due to the limit of resources, only a few sensors know their location. One typical approach to locate sensors is to complete the current incomplete distance matrix via matrix completion. In particular, when the sensor position changes slowly, the previous distance matrix is very similar to the current one, which can be used as prior information.
\end{itemize}
Motivated by the above examples, it is highly desirable to utilize side information to improve the performance of matrix completion.

In this paper, we naturally generalize the recovery procedure \eqref{OurMatrixRecovery} to integrate prior information into matrix completion
\begin{equation}\label{OurMatrixCompletion}
\min_{\vX\in \R^{n \times n}}  \norm{\vX}_{*} - \lambda \ip{\vPhi}{\vX} \quad \text{s.t.}~\vY=\mathcal{R}_p(\vX),
\end{equation}
where $\vY \in \R^{n \times n}$ is the matrix of measurements and $\mathcal{R}_p(\cdot)$ denotes the Bernoulli sampling operator which is defined as
\begin{equation} \label{eq:def of R(Z)}
\mathcal{R}_p(\vX) = \sum_{i,j=1}^{n}\frac{\delta_{ij}}{p_{ij}} \ip{\ve_{i}\ve_{j}^T}{\vX} \ve_{i}\ve_{j}^T.
\end{equation}
Here $\{\delta_{ij}\}$ are i.i.d. Bernoulli random variables which take $1$ with probability $p_{ij}$ and $0$ with probability $1-p_{ij}$. It is not hard to see that  we can observe $m=\sum_{i,j=1}^{n} p_{ij}$ elements in expectation. We then establish performance guarantees for this approach. Specifically, we show that with suitable side information, this approach \eqref{OurMatrixCompletion} can decrease the sample complexity by a logarithmic factor compared with the standard matrix completion procedure. It is worth pointing out that the extension of our theory from matrix recovery \eqref{OurMatrixRecovery} to matrix completion \eqref{OurMatrixCompletion} is not straightforward and requires totally different analytical techniques.

\begin{table*}
	\caption{Summary of different matrix completion approaches with side information. The parameters $\alpha,\beta,$ and $\theta$ are related to the quality of side information.}
	\label{table: comparison}
	\centering
	\begin{tabular}{|c|c|c|c|c|c|}
		\toprule
		Approach & Sample complexity   & Quality of side information & Dimension of subspace side information &  Reference\\
		\midrule	
		MC        &  $O(rn \log^2 n)$  & None      & None  &  \cite{candes2009exact} \\
		IMC       &  $O(rn \log n)$    & Perfect   & $s>r$ &  \cite{jain2013provable} \\
		Dirty IMC &  $O(rn^{3/2} \log n)$          & Imperfect & $s>r$ &  \cite{chiang2015matrix}\\
		DWMC & $O(\beta rn \log^2 n)$                   & Imperfect & $s=r$ &  \cite{chen2015completing}\\
		WMC       &  $O(rn\log n \log\theta n)$                  & Imperfect & $s=r$ &  \cite{eftekhari2018weighted}\\
		Ours      &   $O(rn\log n \log (\alpha n/\log n))$                  & Imperfect & $s=r$ & This paper \\
		\bottomrule
	\end{tabular}
\end{table*}

\subsection{Related Works}
Matrix completion refers to recovering a low-rank matrix from a small number of its random entries. To complete the matrix, the standard way is to solve the following nuclear norm minimization problem \cite{fazel2002matrix,recht2010guaranteed,candes2009exact}
\begin{equation}\label{ClassicalMatrixCompletion}
\min_{\vX} \norm{\vX}_{*} \quad  \text{s.t.}~ \vY = \mathcal{R}_p(\vX).
\end{equation}
The related performance guarantees for \eqref{ClassicalMatrixCompletion} have been extensively studied in the literature, see e.g. \cite{candes2009exact,gross2011recovering,candes2010matrix,candes2010power, chen2015completing,keshavan2010matrix,keshavan2010matrix2,koltchinskii2011nuclear,chandrasekaran2011rank, jain2013low,chen2015incoherence} and references therein. The theoretical results indicate that $O(rn \log^2 n)$ samples are sufficient to accurately complete the matrix for an incoherent rank-$r$ matrix.

Matrix completion with different kinds of prior subspace information has also been studied recently. For instance, with perfect $s$-dimensional subspace information ($s > r$), the Inductive Matrix Completion (IMC) is suggested in \cite{xu2013speedup,jain2013provable} to obtain a better recovery performance than the standard matrix completion procedure. A following work with imperfect $s$-dimensional subspace information, named Dirty IMC, is proposed in \cite{chiang2015matrix}, where the original matrix is completed by splitting it into a low-rank estimate in the subspace and a low-rank perturbation outside the subspace.

Another line of work takes advantage of $r$-dimensional imperfect subspace information to improve the performance of matrix completion.  In \cite{srebro2010collaborative,foygel2011learning,negahban2012restricted,chen2015completing}, the authors propose a diagonal weighted matrix completion (DWMC) method
\begin{equation} \label{DiagonalWeightedMatrixCompletion}
\min_{\vX} \norm{\vR \vX \vC}_* \quad  \text{s.t.}~ \mathcal{P}_\Omega(\vX) =	\mathcal{P}_\Omega(\vX^\star),
\end{equation}
where $\vR=\diag\{r_1,\ldots,r_n\}$ and $\vC=\diag\{c_1,\ldots,c_n\}$ are the diagonal weighted matrices respectively determined by the leverage scores of prior subspace information ${\widetilde{\mathcal{U}}}_r$ and ${\widetilde{\mathcal{V}}}_r$, $\Omega$ is a random index set with
$$
[\mathcal{P}_\Omega(\vX)]_{ij}=
\left\{
{\begin{array}{rl}
	X_{ij}, & (i,j) \in \Omega,  \\
	0, & (i,j) \notin \Omega. \\
	\end{array} }
\right.
$$
The theoretical results presented in \cite{chen2015completing} have shown that by choosing proper weights, the approach \eqref{DiagonalWeightedMatrixCompletion} can outperform the standard low-rank matrix completion procedure.

In \cite{eftekhari2018weighted}, Eftekhari et al. propose a weighted matrix completion method (WMC) with the aid of prior subspace information ${\widetilde{\mathcal{U}}}_r$ and ${\widetilde{\mathcal{V}}}_r$
\begin{equation} \label{WeightedMatrixCompletion}
\min_{\vX} \norm{\vQ_{{\widetilde{\mathcal{U}}}_r,\tau} \cdot \vX \cdot \vQ_{{\widetilde{\mathcal{V}}}_r,\rho}}_{*} \quad  \text{s.t.}~ \vY = \mathcal{R}_p(\vX),
\end{equation}
where $\vQ_{{\widetilde{\mathcal{U}}}_r,\tau}$ and $\vQ_{{\widetilde{\mathcal{V}}}_r,\rho}$ are defined
as
$$
\vQ_{{\widetilde{\mathcal{U}}}_r,\tau}=\tau \cdot \mathcal{P}_{{\widetilde{\mathcal{U}}}_r} +  \mathcal{P}_{{\widetilde{\mathcal{U}}}_r^\bot} \in \R^{n \times n},
$$
and
$$
\vQ_{{\widetilde{\mathcal{V}}}_r,\rho}=\rho \cdot \mathcal{P}_{{\widetilde{\mathcal{V}}}_r} +  \mathcal{P}_{{\widetilde{\mathcal{V}}}_r^\bot} \in \R^{n \times n}.
$$
Here $\tau$ and $\rho$ are some  weights and $\mathcal{P}_{{\widetilde{\mathcal{U}}}_r}$ and $\mathcal{P}_{{\widetilde{\mathcal{U}}}_r^\bot}$ denote the orthogonal projections onto
${\widetilde{\mathcal{U}}}_r$ and ${\widetilde{\mathcal{U}}}_r^\bot$, respectively. $\mathcal{P}_{{\widetilde{\mathcal{V}}}_r}$ and $\mathcal{P}_{{\widetilde{\mathcal{V}}}_r^\bot}$ are defined likewise. Their results have shown that with suitable side information, this approach can decrease the sample complexity by a logarithmic factor compared with the standard procedure.

Table \ref{table: comparison} provides a summary for the above methods. It is not hard to find that when prior information is reliable, our approach can achieve the state-of-the-art performance. In addition, as shown in the simulation, the proposed method \eqref{OurMatrixCompletion} outperforms others for relatively unreliable prior information.

\subsection{Organization}
The paper is organized as follows. We introduce some useful preliminaries in Section \ref{sec: Preliminaries}. Performance guarantees for matrix completion with prior information via maximizing correlation are presented in Section \ref{sec: Performance guarantees}. A practical application where the prior subspace information is available is analyzed in Section \ref{sec: Performance guarantees 2}. Simulations are included in Section \ref{sec: Simulation}, and the conclusion is drawn in Section \ref{sec: Conclusion}. The proofs are postponed to Appendices.

\section{Preliminaries} \label{sec: Preliminaries}
In this section, we provide some helpful notations, definitions and propositions which will be used later.


\subsection{Convex Geometry}

The \emph{subdifferential} of a convex function $g: \R^n \to \R$ at $\vx^\star$ is defined as
\begin{multline*}
	\partial g(\vx^\star) = \{\vu \in \R^n: g(\vx^\star + \vd) \geq g(\vx^\star) + \langle \vu, \vd \rangle~ \\ \textrm{ for all}~\vd \in \R^n \}.
\end{multline*}

Let $\vX^\star=\vU_r \vSigma_r \vV^T_r$ be the compact SVD of the rank-$r$ matrix $\vX^\star$ with $\vU_r \in \R^{n \times r},\vV_r \in \R^{n \times r}$ and $\bm{\Sigma}_r \in \R^{r \times r}$. The subdifferential of  $\norm{\vX^\star}_*$ is given by \cite{watson1992Characterization,lewis2003mathematics}
\begin{multline*}
\partial \norm{\vX^\star}_* =\vU_r \vV_r^T \\+\left\{\vW:\vW^T\vU_r=\bm{0}, \vW\vV_r=\bm{0}, \norm{\vW} \le 1 \right\}.
\end{multline*}

Let the subspace $\mathcal{T}$ be the support of $\vX^\star$ and $\mathcal{T}^{\bot}$ be its orthogonal complement. Then for any matrix $\vX \in \R^{n \times n}$, the orthogonal projection onto $\mathcal{T}$ is
\begin{equation} \label{eq: PT}
	\mathcal{P}_{\mathcal{T}}(\vX)=\vU_r\vU_r^T \vX + \vX \vV_r \vV_r^T- \vU_r \vU_r^T \vX \vV_r \vV_r^T,
\end{equation}
and the orthogonal projection onto $\mathcal{T}^{\bot}$ is
$$\mathcal{P}_{\mathcal{T}^{\bot}}(\vX)=\vX-\mathcal{P}_{\mathcal{T}}(\vX).$$

\subsection{Two Useful Definitions} 
We then review two definitions which are useful for the analysis of matrix completion.

\begin{definition}[Leverage scores] \label{def: ls} Let the thin SVD  for a rank-$r$ matrix $\vX \in \R^{n \times n}$  be $\vU_r \vSigma_r \vV^T_r$ and define $\mathcal{U}_r={\rm{span}}\{\vU_r\}$ and $\mathcal{V}_r={\rm{span}}\{\vV_r\}$. Then the leverage scores $\mu_{i}(\mathcal{U}_r)$ with respect to the $i$-th row of $\vX$, and $\nu_j(\mathcal{V}_r)$ with respect to the $j$-th column of $\vX$ are defined as
	\begin{align}
	\mu_{i}=\mu_{i}(\mathcal{U}_r)&\triangleq\frac{n}{r} \norm{\vU^T_r\ve_{i}}_2^2, ~i=1,2,\ldots,n,\label{def: mu}\\
	\nu_j=\nu_j(\mathcal{V}_r)&\triangleq\frac{n}{r} \norm{\vV^T_r\ve_{j}}_2^2, ~j=1,2,\ldots,n.\label{def: nu}
	\end{align}
\end{definition}

Then the {\it{coherence parameter}} \cite{candes2009exact} of $\vX$ can be expressed as
$$
\eta(\vX)=\max_{i,j}\{\mu_{i}(\mathcal{U}_r),\nu_j(\mathcal{V}_r)\}.
$$
It is not hard to verify that $\eta(\vX) \in [1,\frac{n}{r}]$. Moreover, when $\eta(\vX)$ is small, i.e., $\mathcal{U}_r$ and $\mathcal{V}_r$ are spanned by vectors with nearly equal entries in magnitude, we call that $\vX$ is {\it{incoherent}}; when $\eta(\vX)$ is large, i.e., $\mathcal{U}_r$ or $\mathcal{V}_r$ contains a ``spiky" basis, we call that $\vX$ is {\it{coherent}}.

For convenience, we define the diagonal matrices $\vM=\diag\{\mu_1,\ldots,\mu_n\} $ and  $\vN=\diag\{\nu_1,\ldots,\nu_n\}$.

The following two norms measure the (weighted) largest entry and largest $\ell_2$ norm of the rows or columns of a matrix, respectively.

\begin{definition}[$\mu(\infty)$ norm and $\mu(\infty,2)$ norm,\cite{chen2015completing}]
	For a rank-$r$ matrix $\vX\in\mathbb{R}^{n\times n}$, we set
	\begin{align*}\norm{\vX}_{\mu(\infty)} &=\norm{ \left(\frac{r \vM}{n}\right)^{-\frac{1}{2}} \vX \left(\frac{r \vN}{n}\right)^{-\frac{1}{2}}}_{\infty}\\
	&=\max_{i,j}\sqrt{\frac{n}{\mu_{i}r}}\cdot|\ve^T_i\vX \ve_j|\cdot\sqrt{\frac{n}{\nu_{j}r}},
	\label{eq:mu inf norm}
	\end{align*}
	where $\norm{\vA}_{\infty}$ returns the largest entry of matrix $\vA$ in
	magnitude.
	Moreover, for a rank-$r$ matrix $\vX\in\mathbb{R}^{n\times n}$, we define
	\begin{align*}
	&\norm{\vX}_{\mu(\infty,2)}\\  &=\norm{ \left(\frac{r \vM}{n}\right)^{-\frac{1}{2}} \vX} _{(\infty,2)}\vee \norm{\left(\frac{r \vN}{n}\right)^{-\frac{1}{2}} \vX^{T}} _{(\infty,2)} \\
	&=\left(\max_{i}\sqrt{\frac{n}{\mu_{i}r}}\norm{\vX^T \ve_i}_{2}\right)\vee\left(\max_{j}\sqrt{\frac{n}{\nu_{j}r}}\norm{\vX \ve_{j}}_{2}\right),
	\end{align*}
	where $a \vee b =\max\{a,b\}$ and $\norm{\vX}_{(\infty,2)}$ denotes the largest $\ell_{2}$ norm of the  rows of $\vX$.
\end{definition}


\subsection{Related Results}
For the convenience of comparison, we introduce the following theoretical result for the standard matrix completion procedure.

\begin{proposition}[Theorem 2, \cite{chen2015completing}] \label{prop: StandardMatrixCompletion} Let $\vX^\star \in \R^{n \times n}$ be a rank-$r$ matrix, and $\vY= \mathcal{R}_p(\vX^\star) \in \R^{n \times n}$ denote the matrix of measurements. Let $\mu_i$ and $\nu_{j}$ be the leverage scores as defined in Definition \ref{def: ls}. If 
\begin{equation*}
1 \ge p_{ij} \gtrsim \frac{\left(\mu_{i}+\nu_{j}\right)r\log^2 n}{n},
\end{equation*}
for all $i,j=1,\ldots,n$, then with high probability, $\vX^\star$ is the unique solution for program \eqref{ClassicalMatrixCompletion}.	Here, $f \gtrsim g$ means that $f$ is greater than $g$ up to a universal constant.
\end{proposition}

From the result of Proposition \ref{prop: StandardMatrixCompletion}, we can conclude that for an incoherent rank-$r$ matrix, we need $O(rn \log^2 n)$ samples to accurately complete the matrix.

\section{Performance Guarantees} \label{sec: Performance guarantees}
In this section, we present a theoretical analysis for the proposed procedure \eqref{OurMatrixCompletion}. The results demonstrate that with suitable prior information, the proposed program only requires  $O(rn \log n)$ samples to correctly complete the matrix for incoherent low-rank matrices, which outperforms the standard matrix completion program \eqref{ClassicalMatrixCompletion} and reaches the state-of-the-art performance. The proof of the result is included in Appendices \ref{sec: Proof1} and \ref{sec: Proof2}.

\begin{theorem} \label{thm: main}
Let $\vX^\star=\vU_r \vSigma_r \vV^T_r$ be the thin SVD of the rank-$r$ matrix $\vX^\star \in \R^{n \times n}$ with $\vU_r \in \R^{n \times r},\vV_r \in \R^{n \times r}$ and $\bm{\Sigma}_r \in \R^{r \times r}$. Let $\mu_i$ and $\nu_{j}$ be the leverage scores as defined in Definition \ref{def: ls}.
	If
	\begin{multline*}
	1 \ge p_{ij}\gtrsim \max \left\{\log\(\frac{\alpha_1^2 n}{r\log n}\),1\right\} \cdot \frac{\left(\mu_{i}+\nu_{j}\right)r\log n}{n} \\
	\cdot \max\left\{\(2\xi_1+\xi_2\)^2,1\right\}
	\end{multline*}
	for all $i,j=1,\ldots,n$, and
	$$ \alpha_2 < \frac{15}{16},
	$$
	then with high probability, we can achieve exact recovery of $\vX^\star$  by solving the program \eqref{OurMatrixCompletion}, where
	\begin{align*}
	\alpha_1&=\norm{\vU_r\vV^T_r-\lambda \mathcal{P}_{\mathcal{T}}(\vPhi)}_F,\\
	\alpha_2&=\norm{\lambda \mathcal{P}_{\mathcal{T}^{\bot}}(\vPhi)},\\
	\xi_1&=\norm{\vU_r\vV^T_r-\lambda \mathcal{P}_{\mathcal{T}}(\vPhi)}_{\mu(\infty)},
	\end{align*}
	and
	$$
	\xi_2=\norm{\vU_r\vV^T_r-\lambda \mathcal{P}_{\mathcal{T}}(\vPhi)}_{\mu(\infty,2)}.
	$$	
\end{theorem}

\begin{remark}[No prior information] If there is no prior information, i.e., $\lambda=0$, then the proposed procedure \eqref{OurMatrixCompletion} reduces to the standard one
	\eqref{ClassicalMatrixCompletion}. In this case, simple calculations lead to $\alpha_1=\sqrt{r},\, \alpha_2=0, \, \xi_1  \le 1,$ and $\xi_2=1$. According to Theorem \ref{thm: main}, the bound of sample probability becomes
	\begin{equation*}
	1 \ge p_{ij}\gtrsim \frac{\left(\mu_{i}+\nu_{j}\right)r  \log n \cdot\log\( n/\log n\)}{n}.
	\end{equation*}
	This result implies that for incoherent matrices, $O(rn\log^2 n)$ samples are needed to complete the matrix correctly, which agrees with the result of the standard matrix completion procedure as shown in Proposition \ref{prop: StandardMatrixCompletion}.	
\end{remark}

\begin{remark}[Reliable prior information] \label{rm: optimality} If $\vPhi$ is approximately equal to $\vU_r\vV^T_r$, i.e., $ \mathcal{P}_{\mathcal{T}}(\vPhi)$ is close to $\vU_r\vV^T_r$ and $\norm{\mathcal{P}_{\mathcal{T}^{\bot}}(\vPhi)}$ is small, then we have  $\alpha_1 \to 0$, $\alpha_2 \to 0$, $\xi_1 \to 0$, $\xi_2 \to 0$, and $\log\(\frac{\alpha_1^2 n}{r\log n}\) \lesssim 1$ by setting $\lambda=1$. This means that for incoherent matrices, $m=O(rn\log n)$ samples are sufficient to complete the matrix in this case, which reduces the sample complexity of the standard matrix completion procedure by a logarithmic factor.
\end{remark}

\begin{remark}[Choice of $\vPhi$] Actually, the recovery procedure \eqref{OurMatrixCompletion} provides a general framework for matrix completion with prior information via maximizing correlation. In practice, we need to choose a suitable $\vPhi$ which encodes available prior information effectively. Generally speaking, the choice of $\vPhi$ is problem-specific and usually determined by the prior information in hand. In the subsequent section, we will demonstrate how to choose a suitable $\vPhi$ when the prior subspace information is accessible, and present a theoretical analysis for this application. Another example can be found in \cite{zhang2020}.
\end{remark}
The above main result can be naturally extended the noisy case.

\begin{corollary}
Let $\vX^\star=\vU_r \vSigma_r \vV^T_r$ be the thin SVD of the rank-$r$ matrix $\vX^\star \in \R^{n \times n}$ with $\vU_r \in \R^{n \times r},\vV_r \in \R^{n \times r}$ and $\bm{\Sigma}_r \in \R^{r \times r}$. Consider the noisy observation $\vY=\vX^\star + \vN$, where the entries of the noise $\vN$ is bounded. Let $\mu_i$ and $\nu_{j}$ be the leverage scores as defined in Definition \ref{def: ls}. For the noisy version of matrix completion with prior information via maximizing correlation
\begin{equation}\label{NoisyMatrixCompletion}
\min_{\vX}  \norm{\vX}_{*} - \lambda \ip{\vPhi}{\vX} \quad {\rm{s.t.}}~ \norm{\mathcal{R}_p(\vY-\vX)}_F \le \varepsilon,
\end{equation}
where $\varepsilon$ denotes the upper bound (in term of Frobenius norm) of $\mathcal{R}_p(\vN)$.
If
\begin{multline*}
1 \ge p_{ij}\gtrsim \max \left\{\log\(\frac{\alpha_1^2 n}{r\log n}\),1\right\} \cdot \frac{\left(\mu_{i}+\nu_{j}\right)r\log n}{n} \\
\cdot \max\left\{\(2\xi_1+\xi_2\)^2,1\right\}
\end{multline*}
for all $i,j=1,\ldots,n$, and
$$ \alpha_2 < \frac{7}{8},
$$
then with high probability, by solving the program \eqref{NoisyMatrixCompletion}, the solution $\breve{\vX}$ obeys
\begin{equation*}
\norm{\vX^\star-\breve{\vX}}_{F}\le \[2 + 32 \sqrt{1+\frac{2 n}{r \log n}} \cdot (\sqrt{n}+\norm{\lambda \vPhi}_F) \]\cdot \varepsilon,
\end{equation*}
where
\begin{align*}
\alpha_1&=\norm{\vU_r\vV^T_r-\lambda \mathcal{P}_{\mathcal{T}}(\vPhi)}_F,\\
\alpha_2&=\norm{\lambda \mathcal{P}_{\mathcal{T}^{\bot}}(\vPhi)},\\
\xi_1&=\norm{\vU_r\vV^T_r-\lambda \mathcal{P}_{\mathcal{T}}(\vPhi)}_{\mu(\infty)},
\end{align*}
and
$$
\xi_2=\norm{\vU_r\vV^T_r-\lambda \mathcal{P}_{\mathcal{T}}(\vPhi)}_{\mu(\infty,2)}.
$$	
\end{corollary}

\section{Case Study: prior subspace information} \label{sec: Performance guarantees 2}

In this section, we study a typical example where the desired low-rank matrix is symmetric and its corresponding prior subspace information is available to us. This model has many potential applications in machine learning such as semi-supervised clustering \cite{yi2013semi,bair2013semi} and link prediction \cite{mishra2007clustering,chen2014clustering}.

Let ${\mathcal{U}}_r={\rm span}(\vU_r)$ denote the $r$-dimensional column space of $\vX^\star$ and ${\widetilde{\mathcal{U}}}_r$ be the corresponding prior subspace information.  To leverage the prior subspace information, we modify matrix completion procedure \eqref{OurMatrixCompletion} as follows
\begin{equation}\label{OurMatrixCompletion2}
\min_{\vX\in \R^{n \times n}}  \norm{\vX}_{*} - \lambda \ip{\vPhi}{\vX} \quad \text{s.t.}~\vY=\mathcal{R}_p(\vX),
\end{equation}
where $\vPhi= \widetilde{\vU}_r\widetilde{\vU}_r^T$, the columns of $\widetilde{\vU}_r \in \R^{n \times r}$ constitute the orthonormal bases of the subspace ${\widetilde{\mathcal{U}}}_r $, and $\lambda \in [0,1]$ is the tradeoff parameter. Define the leverage scores of the subspace
$\breve{\mathcal{U}}=\text{\rm{span}}([\vU_r,\widetilde{\vU}_r])$ as follows
\begin{align*}
\breve{\mu}_{i}&\triangleq\mu_{i}(\breve{\mathcal{U}}),~i=1,2,\ldots,n.\label{def: mu_c}
\end{align*}

Then we have the following result. The proof is included in Appendix \ref{Appendix: general case of Thm 2} where an arbitrary low-rank matrix $\vX^\star$ (not limited to symmetric matrices) is considered.

\begin{theorem} \label{thm: main2}
    Let $\vX^\star=\vU_r \vSigma_r \vU^T_r$ be the thin SVD of the rank-$r$ matrix $\vX^\star \in \R^{n \times n}$ with $\vU_r \in \R^{n \times r}$. Denote the column subspace of $\vX^\star$ by ${\mathcal{U}}_r={\rm span}(\vU_r)$ and its corresponding $r$-dimensional prior subspace information by ${\widetilde{\mathcal{U}}}_r$. Define $\vGamma={\rm{diag}}\(\gamma_{1},\ldots,\gamma_{r}\) \in \R^{r \times r}$ whose entries are the principal angles between ${\mathcal{U}}_r$ and ${\widetilde{\mathcal{U}}}_r$. Let $\mu_i$ and $\breve{\mu}_{i}$ be the leverage scores defined as before.
	If
	\begin{multline*}
	1 \ge p_{ij}\gtrsim \max \left\{\log\(\frac{\alpha_1^2 n}{r\log n}\),1\right\} \cdot \frac{ \mu_{i} r\log n}{n} \\
	\cdot \max\left\{\alpha_3^2\beta^2,1\right\}
	\end{multline*}
	for all $i,j=1,\ldots,n$, and
	$$ \alpha_2 < \frac{15}{16},
	$$
	then with high probability, we can achieve exact recovery of $\vX^\star$  by solving the program \eqref{OurMatrixCompletion2}, where
	\begin{align*}
	\alpha_1^2&=\lambda^2 \[r-\sum_{i=1}^{r} \sin^4\gamma_i\]-2 \lambda \sum_{i=1}^{r} \cos^2\gamma_i+r,\\
	\alpha_2&= \lambda \, \max_{i} \{\sin^2 \gamma_i\},\\
	\alpha_3&=\max_i \{1-\lambda \cos^2\gamma_i \}+2\lambda \, \max_i\{\cos\gamma_i \sin \gamma_i\},
	\end{align*}
	and
	$$
	\beta= 1 \vee \sqrt{2 \max_i\frac{\breve{\mu}_{i}}{\mu_{i}}}.
	$$	
\end{theorem}

%

\begin{remark}[Choice of $\lambda$]
    Clearly, the sample complexity is influenced by parameters $\alpha_1, \alpha_2, \alpha_3$, and $\beta$. However, it is not hard to find that $\alpha_1^2$ is the deciding factor. Thus we can choose
    \begin{equation} \label{eq: Optimal_lambda2}
	\lambda^\star=\frac{\sum_{i=1}^{r} \cos^2\gamma_i}{r-\sum_{i=1}^{r} \sin^4\gamma_i}
	\end{equation}
    such that $\alpha_1^2$ achieve its minimum
    $$
	\alpha_1^2 = r- \frac{\[\sum_{i=1}^{r} \cos^2\gamma_i\]^2}{r-\sum_{i=1}^{r} \sin^4\gamma_i},
	$$	
    which will lead to the optimal sample complexity. In particular, when the prior subspace information is close to the original subspace, the best choice of $\lambda$ is $\lambda^\star \approx 1$.
\end{remark}

\begin{figure*}[!t]
	\centering
	\subfloat[]{\includegraphics[width=2.7in]{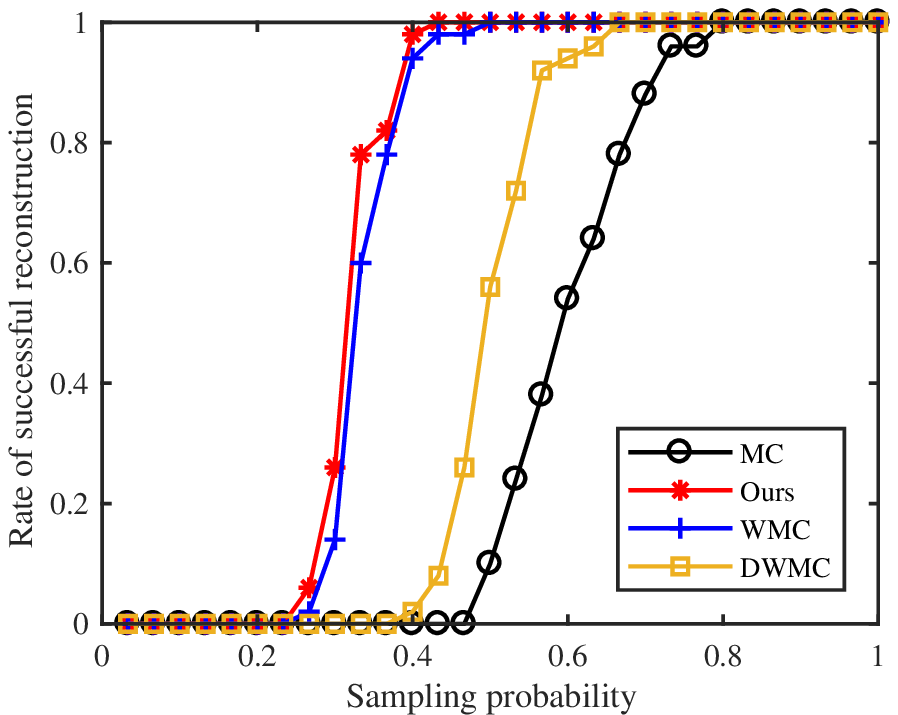}}
	\hfil
	\subfloat[]{\includegraphics[width=2.7in]{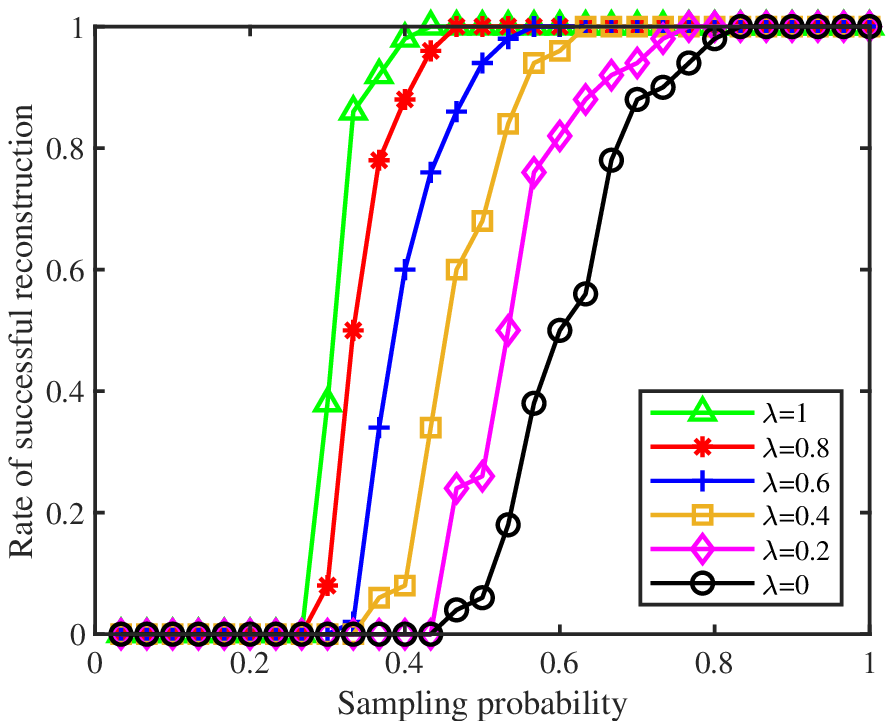}}
	\caption{Rate of successful reconstruction v.s. sampling probability for matrix completion with prior information. (a) Comparisons for MC, our approach, WMC and DWMC. (b) Matrix completion via maximizing correlation with different weights $\lambda$.
	}
	\label{fig: PerformanceComparison_001}
\end{figure*}

\begin{figure*}[!t]
	\centering
	\subfloat[]{\includegraphics[width=2.7in]{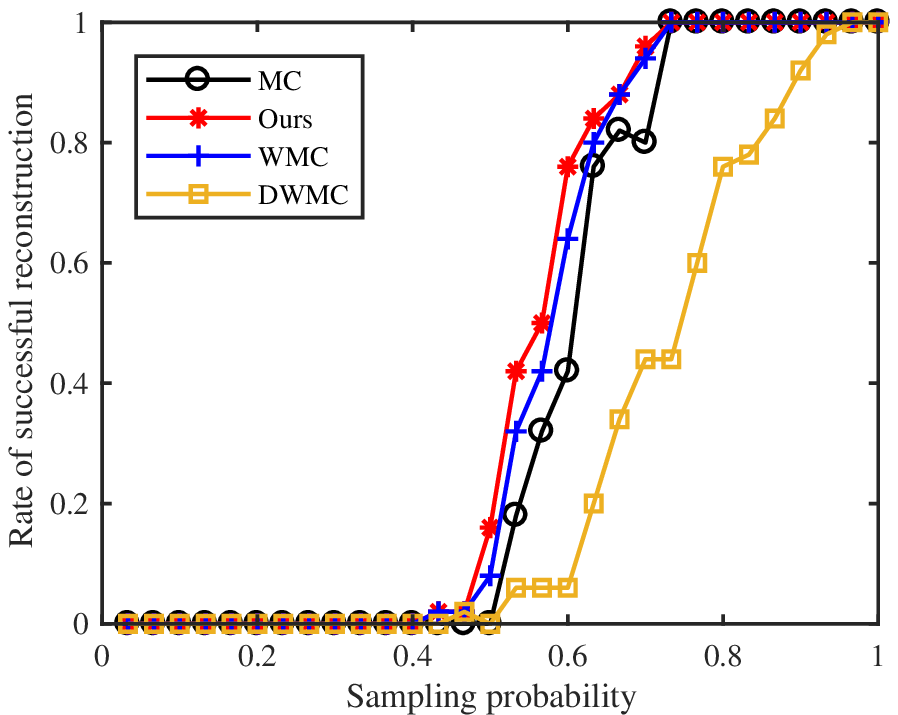}}
	\hfil
	\subfloat[]{\includegraphics[width=2.7in]{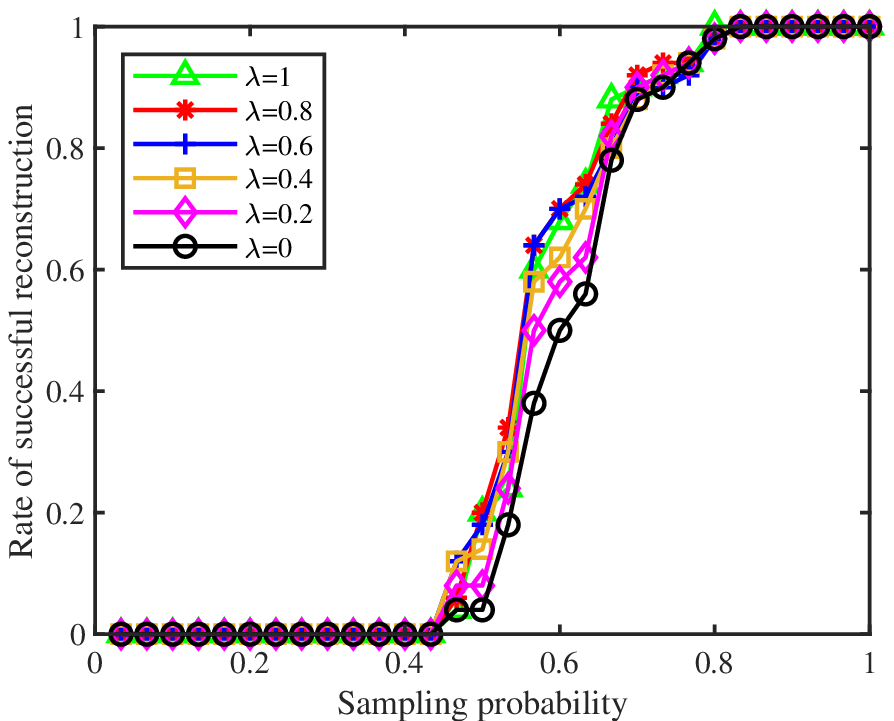}}
	\caption{Rate of successful reconstruction v.s. sampling probability  for matrix completion with weaker prior information. (a) Comparisons for MC, our approach, WMC and DWMC. (b) Matrix completion via maximizing correlation with different weights $\lambda$.}
	\label{fig: PerformanceComparison_01}
\end{figure*}

\begin{figure*}[!t]
	\centering
	\subfloat[]{\includegraphics[width=2.7in]{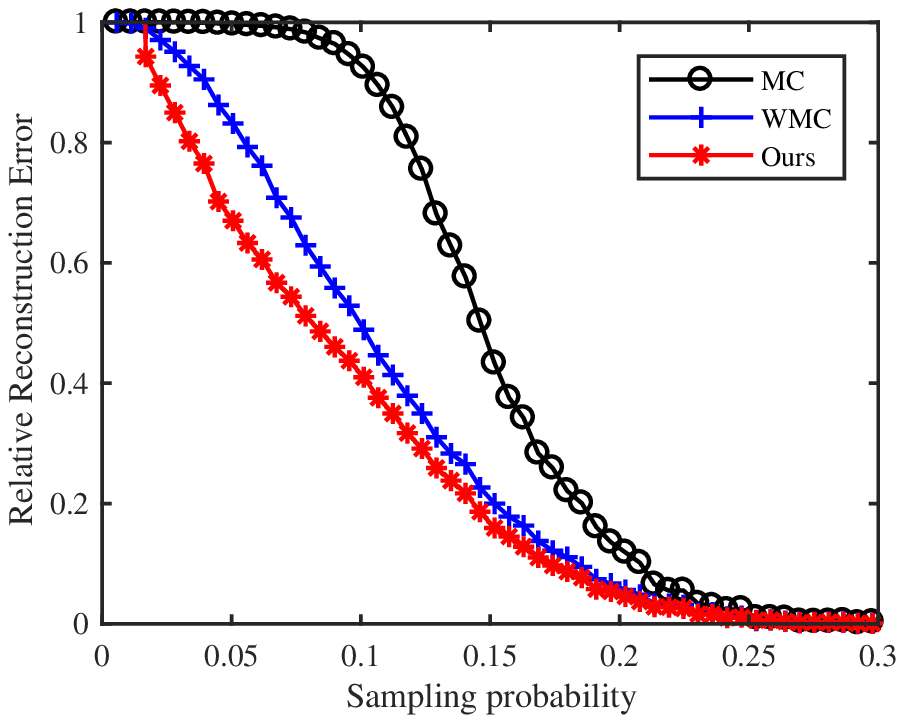}}
	\hfil
	\subfloat[]{\includegraphics[width=2.7in]{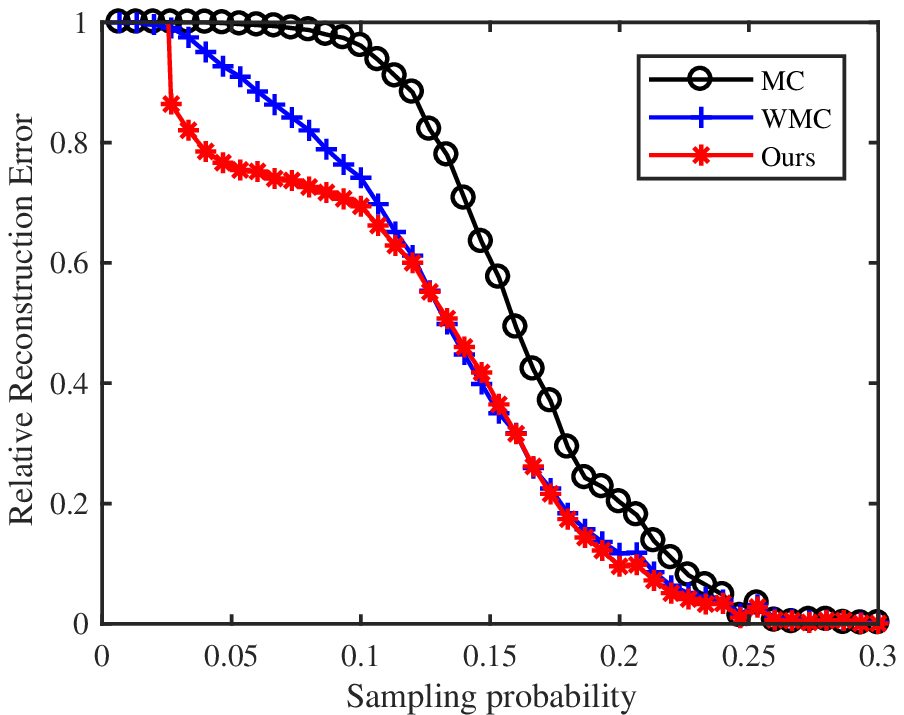}}
	\caption{Relative reconstruction error v.s. sampling probability for different datasets: (a) Wine (b) Iris.}
	\label{fig: PerformanceComparison_03}
\end{figure*}

\section{Simulations and Experiments} \label{sec: Simulation}
In this section, we verify the validity of our theoretical results by both synthetic simulations and real-world applications.

\subsection{Synthetic Simulations}
Let $\vX^\star \in \R^{n \times n}$ be the original rank-$r$ matrix. We construct $\vX^\star$ as follows: generate two independent Gaussian matrices $\vG_1, \vG_2 \in \R^{n \times r}$; let $\vU_r\in\R^{n \times r}$ and $\vV_r\in\R^{n \times r}$ be the basis matrices of subspaces $ {\mathcal{U}}_r={\rm span}\{\vG_1\}$ and ${\mathcal{V}}_r={\rm span}\{\vG_2\}$, respectively; then we construct $\vX^\star = \vU_r \vSigma_r \vV_{r}^T$, where $\vSigma_r=\diag \{\frac{1}{\sqrt{r}},\ldots,\frac{1}{\sqrt{r}}\}$. The prior information is the perturbed matrix $\hat{\vX}=\vX^\star+\sigma \vZ \in \R^{n \times n}$, where the entries of $\vZ$ obey i.i.d. standard normal distribution and $\sigma>0$ is an absolute constant. By taking truncated rank-$r$ SVD for $\hat{\vX}$, we obtain the prior matrix $\vPhi=\hat{\vU}_r \hat{\vV}_r^T$.

We set $n=32$, $r=4$, and $tol=10^{-3}$ in these synthetic experiments. For a specific sampling probability $p$, we make 50 trials, count the number of successful trials, and calculate the related probability. If the solution $\breve{\vX}$ satisfies
$$
\frac{\|\vX^\star - \breve{\vX}\|_\text{F}}{\norm{\vX^\star}_\text{F}} < tol,
$$
then the trial is successful, otherwise it fails. Let $p$ increase from $0$ to $1$ with step $1/n$, we can obtain the simulation results.

We consider two kinds of prior information: standard prior information ($\sigma=0.01$) and weaker prior information ($\sigma=0.1$). For each kind of prior information, we compare the performance of four methods: the standard matrix completion approach \eqref{ClassicalMatrixCompletion}, the proposed procedure \eqref{OurMatrixCompletion}, the diagonal weighted matrix completion \eqref{DiagonalWeightedMatrixCompletion} and the weighted matrix completion \eqref{WeightedMatrixCompletion}
\footnote{Here, we only compare the methods which work with $r$-dimensional subspace information. It seems unfair to compare with  IMC and noisy IMC since they require higher dimensional subspace information ($s > r$).}. For the proposed approach \eqref{OurMatrixCompletion}, $\lambda$ is chosen by following Remark \ref{Choice_of_Lambda_Genaral}. For WMC \eqref{WeightedMatrixCompletion}, we set $w\triangleq\tau=\rho$. As suggested in \cite{eftekhari2018weighted}, the proper choice of $w$ is $w^2=\sqrt{\tan ^{4} \theta+\tan ^{2} \theta}-\tan ^{2} \theta$, where $\theta$ is the largest principal angle. In addition, we run some simulations for \eqref{OurMatrixCompletion} under different weights $\lambda$.

The results of matrix completion under standard prior information ($\sigma=0.01$) are shown in Fig. \ref{fig: PerformanceComparison_001}. Fig.  \ref{fig: PerformanceComparison_001}(a) presents the comparisons for the four methods.  The results illustrate that the proposed approach achieves the best performance. Although the diagonal weighted matrix completion has a worse performance than the proposed approach and the weighted approach, it performs much better than the standard one.  Fig.  \ref{fig: PerformanceComparison_001}(b) shows the performance of the proposed approach under different weights $\lambda$. The results indicate that the integration of side information can reduce the sample complexity of the standard matrix completion ($\lambda=0$). Furthermore, with reliable prior information, the larger the parameter $\lambda$, the better the performance. The optimal $\lambda$ calculated by Eq. \eqref{eq: Optimal_lambda} is $\lambda^\star=0.9895$, which is very close to $1$ and coincides with the simulation results.


In Fig. \ref{fig: PerformanceComparison_01}, we repeat the simulations under weaker prior information ($\sigma=0.1$). In Fig. \ref{fig: PerformanceComparison_01}(a), the performance of the proposed, weighted and diagonal method deteriorates sharply compared with the plots in Fig. \ref{fig: PerformanceComparison_001}(a). The results show the proposed method has the best performance.  We also see that the proposed method and the weighted method slightly outperform the standard matrix completion  while the diagonal method underperforms the standard one. In Fig. \ref{fig: PerformanceComparison_01}(b), all the results for different $\lambda$ almost coincide together, showing a slightly improvement than the standard matrix completion.


\subsection{Real-world Experiments}

Semi-supervised clustering is an important machine learning problem which can be transformed into a low-rank matrix completion problem. Let $\vS$ denote the similarity matrix to be completed, where $S_{ij}=1$ if item $i$ and $j$ are similar, $0$ if dissimilar, and $?$ if similarity is unclear. Let $\vZ$ denote the side information (feature matrix). Our goal is to find a good clustering such that similar items belong to the same cluster, i.e., to label the unknown similarity to promote the low-rankness of the similarity matrix by using the side information.

\begin{table}
	\caption{Statistics of datasets. $\theta_1,\theta_2,$ and $\theta_3$ denote the principal angles between subspaces.}
	\label{table: comparison2}
	\centering
	\begin{tabular}{|c|c|c|c|c|c|c|}
		\toprule
		Dataset & \makecell[c]{No. of \\Classes}   & \makecell[c]{No. of \\items} & \makecell[c]{No. of \\features} &  $\theta_1$ & $\theta_2$& $\theta_3$\\
		\midrule	
		Wine      & 3  & 178  & 13 &     0.13 & 0.44 & 0.65 \\
		Iris      & 3  &150   & 4  &     0.12 & 0.38 & 1.44 \\
		\bottomrule
	\end{tabular}
\end{table}

The experiments are made by using two real-world datasets, wine and iris \cite{Chang2011LIBSVM}. The statistics of the two datasets are shown in Table \ref{table: comparison2}. We compare the performance of three schemes: the standard matrix completion, the proposed method, and the weighted matrix completion. Augmented Lagrange Multiplier (ALM) method is used to solve the models \cite{lin2010augmented}. The subspace information $\hat{\vU}_r$ is extracted from the feature matrix, where $\hat{\vU}_r$ is the left singular matrix generated from the truncated rank-$r$ SVD of $\vZ$. We set $\hat{\vV}_r=\hat{\vU}_r$ since the similar matrix $\vS$ are symmetric. The samples are chosen randomly and symmetrically. For each sampling probability, we make $50$ trials and calculate the average of relative reconstruction error $\|\vX^\star - \breve{\vX}\|_\text{F}/\norm{\vX^\star}_\text{F}$, where $\breve{\vX}$ denotes the recovered solution.

The results are presented in Fig. \ref{fig: PerformanceComparison_03}. The parameter $\lambda$ is chosen according to Eq. \eqref{eq: Optimal_lambda2} and $w$ is set as before.  Both Fig. \ref{fig: PerformanceComparison_03}(a) and Fig. \ref{fig: PerformanceComparison_03}(b) show that the proposed method achieves the best performance and WMC shows a better performance than the standard MC. The results illustrate that our procedure seems more robust than the weighed method in this application.


\section{Conclusion} \label{sec: Conclusion}

In this paper, we have suggest a strategy to complete a low-rank matrix from a small collections of its random entries with the aid of prior information. We have established the performance guarantees for the proposed method. The results have illustrated that with reliable side information, the proposed method can decrease the number of measurements of the standard matrix completion procedure by a logarithmic factor. We also have presented a typical example in which the prior subspace information is available to us. The choice of the tradeoff parameter has also been considered. Numerical experiments have been provided to verify the theoretical results.

In terms of future work, it is worthwhile to study how to choose the suitable function $\vPhi$ in Eq. \eqref{OurMatrixCompletion} based on the available prior information to improve the performance of the proposed approach. Besides, it is interesting to present some theoretical insights why the proposed approach has more robust performance than the weighted algorithm.

\appendices

\section{Proof of Theorem \ref{thm: main}} \label{sec: Proof1}

Before proving Theorem \ref{thm: main}, we require some lemmas, in which Lemma \ref{lm: DualCertificate} is proved in Section \ref{sec: Proof2}.

\begin{lemma}[Lemma 9, \cite{chen2015completing}]
	\label{lemma:near isometry 1}  For probabilities $\{p_{ij}\}\subset(0,1]$, consider
	the measurement operator $\mathcal{R}_p(\cdot)$ defined in \eqref{eq:def of R(Z)} and projection operator $\mathcal{P}_{\mathcal{T}}$  defined in \eqref{eq: PT}. Then, except with a probability of at most $n^{-20}$,
	$$
	\left\Vert \left(\mathcal{P}_{\mathcal{T}}-\mathcal{P}_{\mathcal{T}}\mathcal{R}_p\mathcal{P}_{\mathcal{T}}\right)(\cdot)\right\Vert _{F\rightarrow F}\le\frac{1}{2},
	$$
	provided that
	\begin{equation}\label{eq:no of samples}
	\frac{\left(\mu_{i}+\nu_{j}\right)r\log n}{n} \lesssim p_{ij} \le 1
	,
	\qquad \forall i,j\in[1:n],
	\end{equation}
	where $\|\mathcal{A}(\cdot)\|_{F\rightarrow F}=\sup_{\|\vX\|_{F}\le1}\|\mathcal{A}(\vX)\|_{F}$
	, and $(\mathcal{A}  \mathcal{B})(\cdot) = \mathcal{A}(\mathcal{B}(\cdot))$.
\end{lemma}

\begin{lemma}[Lemma 13, \cite{chen2015completing}] \label{lm: chen_lemma13}
	If  the projection operator $\mathcal{P}_{\mathcal{T}}$ satisfies $
	\left\Vert \left(\mathcal{P}_{\mathcal{T}}-\mathcal{P}_{\mathcal{T}}\mathcal{R}_p\mathcal{P}_{\mathcal{T}}\right)(\cdot)\right\Vert _{F\rightarrow F}\le\frac{1}{2},
	$
	then we have
	\begin{multline*}
	\norm{ \mathcal{P}_{\mathcal{T}}(\vZ)}_{F}\le  \(\max_{i,j} \sqrt{\frac{2}{p_{ij}}}\)\norm{\mathcal{P}_{\mathcal{T}^{\bot}}(\vZ)}_F, \\ \forall\, \vZ \in \{\vZ:\mathcal{R}_p(\vZ)=0\}.
	\end{multline*}
\end{lemma}

\begin{lemma} \label{lm: DualCertificate}
Let $\vX^\star=\vU_r \vSigma_r \vV^T_r$ be the compact SVD of the rank-$r$ matrix $\vX^\star$. Let the subspace $\mathcal{T}$ be the support of $\vX^\star$ and $\mathcal{T}^{\bot}$ be its orthogonal complement. Let $\mu_i$ and $\nu_{j}$ be the leverage scores as defined in Definition \ref{def: ls}. Let $l^{-1}$ is polynomial in $n$.
If
\begin{multline*}
1 \ge p_{ij}\gtrsim \max \left\{\log\(\frac{\alpha_1}{l}\),1\right\} \cdot \frac{\left(\mu_{i}+\nu_{j}\right)r\log n}{n} \\
\cdot \max\left\{\(2\xi_1+\xi_2\)^2,1\right\}
\end{multline*}
for all $i,j=1,\ldots,n$, and
\begin{equation*}
	\alpha_2 < \frac{15}{16},
\end{equation*}
then with high probability, there exists $\vY \in {\rm range} (\mathcal{R}_p)$ satisfying
$$\norm{\lambda \mathcal{P}_{\mathcal{T}^{\bot}}(\vPhi)+\mathcal{P}_{\mathcal{T}^{\bot}}(\vY)} < \frac{1}{32}+\alpha_2,$$
and
$$
\norm{\vU_r\vV^T_r -\lambda \mathcal{P}_{\mathcal{T}}(\vPhi)- \mathcal{P}_{\mathcal{T}}(\vY)}_F \le \frac{l}{32\sqrt{2}},
$$
where
\begin{align*}
\alpha_1&=\norm{\vU_r\vV^T_r-\lambda \mathcal{P}_{\mathcal{T}}(\vPhi)}_F,~
\alpha_2=\norm{\lambda \mathcal{P}_{\mathcal{T}^{\bot}}(\vPhi)},\\
\xi_1&=\norm{\vU_r\vV^T_r-\lambda \mathcal{P}_{\mathcal{T}}(\vPhi)}_{\mu(\infty)},
\end{align*}
and
$$
\xi_2=\norm{\vU_r\vV^T_r-\lambda \mathcal{P}_{\mathcal{T}}(\vPhi)}_{\mu(\infty,2)}.
$$	
\end{lemma}

Now we are ready to prove Theorem \ref{thm: main}. Consider any feasible solution ($\vX^\star +\vZ$) to problem \eqref{OurMatrixCompletion} for non-zero matrix $\vZ \in \{\vZ:\mathcal{R}_p(\vZ)=0\}$. Let $\vW \in \R^{n \times n}$ be a matrix satisfying $\vW \in \left\{\vW:\vW^T\vU_r=\bm{0}, \vW\vV_r=\bm{0}, \norm{\vW} \le 1 \right\}$ and $\ip{\vW}{\mathcal{P}_{\mathcal{T}^{\bot}}(\vZ)}=\norm{\mathcal{P}_{\mathcal{T}^{\bot}}(\vZ)}_*$. Then we have $\vW = \mathcal{P}_{\mathcal{T}^{\bot}}(\vW)$ and $\vU_r\vV_r^T+\vW \in \partial \norm{\vX^\star}_*$. According to the definition of subdifferential, for any non-zero matrix $\vZ \in {\rm{ker}}(\mathcal{R}_p)$, we have
\begin{multline} \label{neq: subdiff}
\norm{\vX^\star+\vZ}_{*}-\lambda \ip{\vPhi}{\vX^\star+\vZ} \\ \ge \norm{\vX^\star}_{*}-\lambda \ip{\vPhi}{\vX^\star}
+\ip{\vU_r\vV_r^T+\vW-\lambda \vPhi}{\vZ}
\end{multline}
Let $\vY \in {\rm{range}}(\mathcal{R}_p)$, then we have $\ip{\vY}{\vZ}=0$ and
\begin{multline*}
\norm{\vX^\star+\vZ}_{*}-\lambda \ip{\vPhi}{\vX^\star+\vZ}
\\ \ge \norm{\vX^\star}_{*}-\lambda \ip{\vPhi}{\vX^\star} +\ip{\vU_r\vV_r^T+\vW-\lambda \vPhi-\vY}{\vZ} \nonumber
\end{multline*}
Using Holder's inequality and the properties of $\vW$ yields
\begin{align*}
&\ip{\vU_r\vV_r^T+\vW-\lambda \vPhi-\vY}{\vZ}\\
& =\ip{\vU_r\vV_r^T-\lambda \mathcal{P}_{\mathcal{T}}(\vPhi)-\mathcal{P}_{\mathcal{T}}(\vY)}{\mathcal{P}_{\mathcal{T}}(\vZ)}\\
&\quad+
\ip{\vW-\lambda \mathcal{P}_{\mathcal{T}^{\bot}}(\vPhi)-\mathcal{P}_{\mathcal{T}^{\bot}}(\vY)}{\mathcal{P}_{\mathcal{T}^{\bot}}(\vZ)}\\
& \ge - \norm{\vU_r\vV_r^T -\lambda \mathcal{P}_{\mathcal{T}}(\vPhi)-\mathcal{P}_{\mathcal{T}}(\vY)}_F \norm{\mathcal{P}_{\mathcal{T}}(\vZ)}_F\\
&\quad+(1-\norm{\lambda \mathcal{P}_{\mathcal{T}^{\bot}}(\vPhi)+\mathcal{P}_{\mathcal{T}^{\bot}}(\vY)})\norm{\mathcal{P}_{\mathcal{T}^{\bot}}(\vZ)}_*\\
& \ge - \norm{\vU_r\vV_r^T -\lambda \mathcal{P}_{\mathcal{T}}(\vPhi)-\mathcal{P}_{\mathcal{T}}(\vY)}_F \norm{\mathcal{P}_{\mathcal{T}}(\vZ)}_F\\
&\quad+(1-\norm{\lambda \mathcal{P}_{\mathcal{T}^{\bot}}(\vPhi)+\mathcal{P}_{\mathcal{T}^{\bot}}(\vY)})\norm{\mathcal{P}_{\mathcal{T}^{\bot}}(\vZ)}_F.
\end{align*}
where the second inequality follows from $\norm{\mathcal{P}_{\mathcal{T}^{\bot}}(\vZ)}_* \ge \norm{\mathcal{P}_{\mathcal{T}^{\bot}}(\vZ)}_F$.

Suppose the assumptions of Lemma \ref{lm: DualCertificate} is satisfied,
using Lemma \ref{lm: DualCertificate} yields
\begin{align} \label{neq: DualResults}
&\ip{\vU_r\vV_r^T+\vW-\lambda \vPhi-\vY}{\vZ} \nonumber\\
&\qquad \ge - \frac{l}{32\sqrt{2}} \norm{\mathcal{P}_{\mathcal{T}}(\vZ)}_F + \(\frac{31}{32}-\alpha_2\)\norm{\mathcal{P}_{\mathcal{T}^{\bot}}(\vZ)}_F \\
&\qquad > - \frac{l}{32\sqrt{2}} \norm{\mathcal{P}_{\mathcal{T}}(\vZ)}_F + \frac{1}{32}\norm{\mathcal{P}_{\mathcal{T}^{\bot}}(\vZ)}_F \ge 0,
\end{align}
where the last inequality applies Lemma \ref{lm: chen_lemma13}, and $\min_{i,j} p_{ij} \ge l^2$. Here, we assign
\begin{equation*}
	l^2\triangleq \frac{r \log n}{n},
\end{equation*}
and the corresponding bound of probability becomes
\begin{multline*}
1 \ge p_{ij}\gtrsim \max \left\{\log\(\frac{\alpha_1^2 n}{r\log n}\),1\right\} \cdot \frac{\left(\mu_{i}+\nu_{j}\right)r\log n}{n} \\
\cdot \max\left\{\(2\xi_1+\xi_2\)^2,1\right\}
\end{multline*}

By incorporating \eqref{neq: DualResults} into \eqref{neq: subdiff}, we have
\begin{equation*}
\norm{\vX^\star+\vZ}_{*}-\lambda \ip{\vPhi}{\vX^\star+\vZ} > \norm{\vX^\star}_{*}-\lambda \ip{\vPhi}{\vX^\star}
\end{equation*}
for any non-zero matrix $\vZ \in \{\vZ:\mathcal{R}_p(\vZ)=0\}$, which completes the proof.

\section{Proof of Lemma \ref{lm: DualCertificate}} \label{sec: Proof2}
In this section, we use the golfing scheme to construct the dual certificate by following \cite{gross2011recovering,chen2015completing,eftekhari2018weighted}. Before proving Lemma \ref{lm: DualCertificate}, let's review some useful lemmas which will be used in the proof.

\begin{lemma} [Lemma 10, \cite{chen2015completing}]
	\label{lemma:any Z} Consider a fixed $\vX \in \R^{n \times n}$. For some universal constant $\Delta \ge 1$,  if
	\begin{equation*}
	\frac{\Delta^2 \left(\mu_{i}+\nu_{j}\right)r\log n}{n} \lesssim p_{ij} \le 1,
	\qquad \forall \, i,j\in[1:n],
	\end{equation*}
	holds, then
	$$
	\left\Vert \left(\mathcal{R}_p-\mathcal{I}\right)(\vX)\right\Vert \le \frac{1}{\Delta}\(\|\vX\|_{\mu(\infty)}+\|\vX\|_{\mu(\infty,2)}\),
	$$
	except with a probability of at most $n^{-20}$. Here, $\mathcal{I}(\cdot)$ is the identity operator.
\end{lemma}

\begin{lemma}[Lemma 11, \cite{chen2015completing}]
	\label{lemma:inf two bnd} Consider  a fixed matrix $\vX\in \mathcal{T}\subset \mathbb{R}^{n\times n}$ (i.e., $\mathcal{P}_{\mathcal{T}}(\vX)=\vX$). Then except with a probability of at most $n^{-20}$, it holds that
	\begin{multline*}
	\norm{\left(\mathcal{P}_{\mathcal{T}}-\mathcal{P}_{\mathcal{T}}\mathcal{R}_p\mathcal{P}_{\mathcal{T}}\right)(\vX)} _{\mu(\infty,2)}  \\ \le \frac{1}{2}\|\vX\|_{\mu(\infty)}+\frac{1}{2}\|\vX\|_{\mu(\infty,2)},
	\end{multline*}
	as long as (\ref{eq:no of samples}) holds.
\end{lemma}

\begin{lemma} [Lemma 12, \cite{chen2015completing}]
	\label{lemma:inf bound}  Consider a fixed matrix $\vX\in T\subset\mathbb{R}^{n\times n}$. Then except with a probability of at most $n^{-20}$, it holds that
	$$
	\left\Vert \left(\mathcal{P}_{\mathcal{T}}-\mathcal{P}_{\mathcal{T}}\mathcal{R}_p\mathcal{P}_{\mathcal{T}}\right)(\vX)\right\Vert _{\mu(\infty)}\le\frac{1}{2}\|\vX\|_{\mu(\infty)},
	$$
	as long as (\ref{eq:no of samples}) holds.
\end{lemma}

Armed with these lemmas, we are ready to prove Lemma \ref{lm: DualCertificate}. In order to measure $\vX^\star$, we use $K$ independent measurement operator $\mathcal{R}_q(\cdot)$ instead of $\mathcal{R}_p(\cdot)$, which means the probability $p_{ij}$ and $q_{ij}$ satisfies
\begin{equation}\label{eq: p_q}
	(1-q_{ij})^K=1-p_{ij},~i,j=1,\ldots,n,
\end{equation}
for given $K$.

Define $\vW_0 = \vU_r\vV^T_r-\lambda \mathcal{P}_{\mathcal{T}}(\vPhi) $ and set $\vY_k = \sum_{j=1}^k \mathcal{R}_{q}(\vW_{j-1})$, $\vW_k = \vU_r\vV^T_r -\lambda \mathcal{P}_{\mathcal{T}}(\vPhi)- \mathcal{P}_{\mathcal{T}}(\vY_k)$ for $k=1,\ldots, K$.  Then for $k=1,\ldots, K$, we have
\begin{align*}
  \vW_k&= \vU_r\vV^T_r -\lambda \mathcal{P}_{\mathcal{T}}(\vPhi)- \mathcal{P}_{\mathcal{T}}(\vY_{k-1}+\mathcal{R}_{q}(\vW_{k-1}))\\
  &= \vU_r\vV^T_r -\lambda \mathcal{P}_{\mathcal{T}}(\vPhi)- \mathcal{P}_{\mathcal{T}}(\vY_{k-1})- \mathcal{P}_{\mathcal{T}}(\mathcal{R}_{q}(\vW_{k-1}))\\
  &=\vW_{k-1}-\mathcal{P}_{\mathcal{T}}(\mathcal{R}_{q}(\vW_{k-1}))\\
  &=\(\mathcal{P}_{\mathcal{T}}-\mathcal{P}_{\mathcal{T}}\mathcal{R}_{q}\mathcal{P}_{\mathcal{T}}\)(\vW_{k-1}).
\end{align*}

According to Lemma \ref{lemma:near isometry 1}, we have
\begin{equation*}
  \norm{\vW_k}_F=\norm{\(\mathcal{P}_{\mathcal{T}}-\mathcal{P}_{\mathcal{T}}\mathcal{R}_{q}\mathcal{P}_{\mathcal{T}}\)(\vW_{k-1})}_F
  \le \frac{1}{2}\norm{\vW_{k-1}}_F,
\end{equation*}
except with a probability of at most $n^{-20}$, as long as
\begin{equation*}
\frac{\left(\mu_{i}+\nu_{j}\right)r\log n}{n} \lesssim q_{ij} \le 1
,
\qquad \forall i,j\in[1:n].
\end{equation*}

By iteration, we obtain
\begin{equation*}
  \norm{\vW_K}_F \le 2^{-K}\norm{\vW_{0}}_F,
\end{equation*}
except with a probability of at most $K n^{-20}$,

Let $\vY=\vY_{K}$, then
\begin{align*}
  \norm{\vW_{K}}_F&=\norm{\vU_r\vV^T_r -\lambda \mathcal{P}_{\mathcal{T}}(\vPhi)- \mathcal{P}_{\mathcal{T}}(\vY)}_F \\
  &\le 2^{-K}\norm{\vW_{0}}_F,
\end{align*}
except with a probability of at most $K n^{-20}$. Let
$$K = \max\left\{\log\(\frac{32 \sqrt{2}\alpha_1}{l}\),1\right\},$$ where $l^{-1}$ is polynomial in $n$. Then we have
\begin{align*}
	\norm{\vU_r\vV^T_r -\lambda \mathcal{P}_{\mathcal{T}}(\vPhi)- \mathcal{P}_{\mathcal{T}}(\vY)}_F &\le
	2^{-K}\norm{\vW_{0}}_F \\
	&= 2^{-K} \alpha_1 \le \frac{l}{32\sqrt{2}},	
\end{align*}
except with a probability of at most
$$K n^{-20}=O(\log(\alpha_1 n)) \cdot n^{-20}=o(n^{-19}).$$

From the triangle inequality, we have
\begin{equation}
  \norm{\lambda \mathcal{P}_{\mathcal{T}^{\bot}}(\vPhi)+\mathcal{P}_{\mathcal{T}^{\bot}}(\vY)}
  \le \norm{\mathcal{P}_{\mathcal{T}^{\bot}}(\vY)} + \norm{\lambda\mathcal{P}_{\mathcal{T}^{\bot}}(\vPhi)}.
\end{equation}
From Lemma \ref{lemma:any Z}, except with a probability of at most
$K n^{-20}=o(n^{-19})$, we have
\begin{align*}
\norm{\mathcal{P}_{\mathcal{T}^{\bot}}(\vY)}& \le \sum_{j=1}^{K}\norm{ \mathcal{P}_{\mathcal{T}^{\bot}} \mathcal{R}_{q}(\vW_{j-1})}\\
& =   \sum_{j=1}^{K}\norm{\mathcal{P}_{\mathcal{T}^{\bot}}\( \mathcal{R}_{q}(\vW_{j-1})-\vW_{j-1}\)}\\
& \le   \sum_{j=1}^{K}\norm{\( \mathcal{R}_{q}-\mathcal{I}\)\vW_{j-1}}\\
& \le\frac{1}{\Delta} \sum_{j=1}^{K}\(\|\vW_{j-1}\|_{\mu(\infty)}+\|\vW_{j-1}\|_{\mu(\infty,2)}\),
\end{align*}
as long as
\begin{equation*}
\frac{ \Delta^2  \left(\mu_{i}+\nu_{j}\right)r\log n}{n} \lesssim q_{ij} \le 1,
\qquad \forall i,j\in[1:n].
\end{equation*}
The second line  holds since $\vW_{j-1}= \mathcal{P}_\mathcal{T} (\vW_{j-1})$ and the third line holds since $\norm{\mathcal{P}_{\mathcal{T}^{\bot}} (\vX)} \le \norm{\vX}$ for any $\vX \in \R^{n \times n}$.
Using Lemma \ref{lemma:inf bound} leads to
\begin{align*}
\norm{\vW_{j-1}}_{\mu(\infty)}&=\norm{\(\mathcal{P}_{\mathcal{T}}-\mathcal{P}_{\mathcal{T}}\mathcal{R}_{q}\mathcal{P}_{\mathcal{T}}\)(\vW_{j-2})}_{\mu(\infty)}\\
&\le \frac{1}{2}\norm{\vW_{j-2}}_{\mu(\infty)},
\end{align*}
except with a probability of at most $n^{-20}$, as long as
\begin{equation*}
\frac{\left(\mu_{i}+\nu_{j}\right)r\log n}{n} \lesssim q_{ij} \le 1
,
\qquad \forall i,j\in[1:n].
\end{equation*}
By iteration, we obtain
$$
\norm{\vW_{j-1}}_{\mu(\infty)} \le \frac{1}{2^{{j-1}}}\norm{\vW_{0}}_{\mu(\infty)}
$$
except with a probability of at most $o(n^{-19})$, since $j \le K$.
By using Lemma \ref{lemma:inf two bnd}, we obtain
\begin{align*}
	\|\vW_{j-1}\|_{\mu(\infty,2)} &=
	\norm{\left(\mathcal{P}_{\mathcal{T}}-\mathcal{P}_{\mathcal{T}}\mathcal{R}_q\mathcal{P}_{\mathcal{T}}\right)(\vW_{j-2})} _{\mu(\infty,2)} \\ &\le\frac{1}{2}\|\vW_{j-2}\|_{\mu(\infty)}+\frac{1}{2}\|\vW_{j-2}\|_{\mu(\infty,2)},\\
	&\le\frac{j-1}{2^{j-1}}\|\vW_{0}\|_{\mu(\infty)}+\frac{1}{2^{j-1}}\|\vW_{0}\|_{\mu(\infty,2)},
\end{align*}
except with a probability of at most $o(n^{-19})$ due to the fact $j \le K$.

It follows that 
\begin{align*}
\norm{\mathcal{P}_{\mathcal{T}^{\bot}}(\vY)} &\le
\frac{1}{\Delta} \sum_{j=1}^{K}\frac{j}{ 2^{j-1}}\norm{\vW_0}_{\mu(\infty)} \\
&\quad + \frac{1}{\Delta} \sum_{j=1}^{K} \frac{1}{2^{j-1}}\norm{\vW_0}_{\mu(\infty,2)}\\
&< \frac{1}{\Delta} \(4\norm{\vW_0}_{\mu(\infty)} +2\norm{\vW_0}_{\mu(\infty,2)}\),
\end{align*}
where $\sum_{j=1}^{K}j\cdot 2^{-(j-1)} < 4$ and $\sum_{j=1}^{K} 2^{-(j-1)} < 2$ for finite $K$.

Then except with a probability of at most $o(n^{-19})$,
\begin{equation*}
\norm{\mathcal{P}_{\mathcal{T}^{\bot}}(\vY)}
< \frac{4 \xi_1+2 \xi_2}{\Delta} = \frac{1}{32},
\end{equation*}
provided that
\begin{equation*}
 \frac{\Delta^2 \left(\mu_{i}+\nu_{j}\right)r\log n}{n} \lesssim q_{ij} \le 1,
\qquad \forall i,j\in[1:n].
\end{equation*}
where we set $\Delta=128 \xi_1+64 \xi_2.$

So we conclude that if
$$
\max \left\{ \(2\xi_1+\xi_2\)^2, 1 \right\} \cdot \frac{\left(\mu_{i}+\nu_{j}\right)r\log n}{n} \lesssim q_{ij} \le 1,
$$
we have
\begin{equation*}
\norm{\lambda \mathcal{P}_{\mathcal{T}^{\bot}}(\vPhi)+\mathcal{P}_{\mathcal{T}^{\bot}}(\vY)} < \frac{1}{32}+\alpha_2
\end{equation*}
and
\begin{equation*}
\norm{\vU_r\vV^T_r -\lambda \mathcal{P}_{\mathcal{T}}(\vPhi)- \mathcal{P}_{\mathcal{T}}(\vY)}_F \le \frac{l}{32\sqrt{2}},	
\end{equation*}
except with a probability of at most $o(n^{-19})$.
Finally, according to \eqref{eq: p_q}, if $\{q_{ij}\}$ are small enough, which means $n$ is large enough, we have
\begin{align*}
	p_{ij}&=1-(1-q_{ij})^K  \gtrsim K q_{ij}\\
	&\gtrsim \max \left\{\log\(\frac{\alpha_1}{l}\),1\right\} \cdot \frac{\left(\mu_{i}+\nu_{j}\right)r\log n}{n}\cdot\\
	&\quad \max\left\{\(2\xi_1+\xi_2\)^2,1\right\}.		
\end{align*}

\section{Proof of a general case of Theorem \ref{thm: main2}} \label{Appendix: general case of Thm 2}

Here, we consider a non-symmetric desired matrix $\vX^\star$.
Let ${\mathcal{U}}_r={\rm span}(\vU_r)$ and ${\mathcal{V}}_r={\rm span}(\vV_r)$ denote the $r$-dimensional column and row spaces of $\vX^\star$, respectively.  Assume that the prior subspace information of the $r$-dimensional column and row spaces of $\vX^\star$, denoted by ${\widetilde{\mathcal{U}}}_r$ and ${\widetilde{\mathcal{V}}}_r$ respectively, is available to us.  By leveraging the prior subspace information, we modify matrix completion procedure \eqref{OurMatrixCompletion} as follows
\begin{equation}\label{OurMatrixCompletion2_general}
\min_{\vX}  \norm{\vX}_{*} - \lambda \ip{\vPhi}{\vX} \quad \text{s.t.}~\vY=\mathcal{R}_p(\vX),
\end{equation}
where $\vPhi= \widetilde{\vU}_r\widetilde{\vV}_r^T$, $\widetilde{\vU}_r \in \R^{n \times r}$ and $\widetilde{\vV}_r \in \R^{n \times r}$ are the orthonormal bases for subspaces ${\widetilde{\mathcal{U}}}_r $ and ${\widetilde{\mathcal{V}}}_r$ respectively, and $\lambda \in [0,1]$ is a tradeoff parameter.

We first introduce an important result in matrix analysis \cite{golub2012matrix,eftekhari2018weighted}. Throughout the paper,
$\vI_n \in \R^{n \times n}$ denotes identity matrix and $\bm{0}_n \in \R^{n \times n}$ denote all-zero matrix. A simple extension of \cite[Lemma 3]{eftekhari2018weighted} achieves the following general result.

\begin{lemma} \label{lem:canonical}
	Consider a rank-$r$ matrix $\vX \in \R^{n \times n}$. Let $\vU_r$ and $\widetilde{\vU}_r \in \R^{n \times r}$ be orthonormal bases for $r$-dimensional subspaces ${\mathcal{U}}_r={\rm{span}}(\vX)$ and ${\widetilde{\mathcal{U}}}_r$, respectively. And let the SVD of  $\vU_r^T \widetilde{\vU}_r= \vL_L \cos (\vGamma) \vR_L^T$, where $\vL_L\in \R^{r \times r}$ and  $\vR_L \in \R^{r \times r}$ are orthogonal matrices, $\vGamma = \diag\{\gamma_1,\ldots,\gamma_r\}\in\mathbb{R}^{r\times r}$ is a diagonal matrix, which contains the principal angles between ${\mathcal{U}}_{r}$ and $\widetilde{\mathcal{U}}_{r}$ with $\pi/2 \ge \gamma_{1}\ge \gamma_{2}\ge\cdots\ge \gamma_{r}\ge 0$. The diagonal matrix $\cos(\vGamma)$ is defined as
	$$
	\cos (\vGamma) \triangleq
	\diag\{\cos \gamma_1, \, \cos \gamma_2, \,\ldots, \, \cos \gamma_r \}
	\in\mathbb{R}^{r\times r},
	$$
	and $\sin(\vGamma)\in\mathbb{R}^{r\times r}$ is defined likewise.
	Then, there exist  $\vU_r',\,\widetilde{\vU}_r'\in \mathbb{R}^{n\times r}$, and $\vU_{n-2r}''\in\mathbb{R}^{n\times (n-2r)}$ such that
	\begin{align*} 	
	\vB_L& =\left[\begin{array}{ccc}
	\vU_{r} & \vU'_{r} & \vU''_{n-2r}\end{array}\right]  \underbrace{ \left[
		\begin{array}{ccc}
		\vL_L   & \\
		& \vL_L \\
		&  & \vI_{n-2r}
		\end{array}
		\right]}_{\triangleq\vC_L} \in\mathbb{R}^{n\times n},
	\nonumber\\
	\widetilde{\vB}_L &=\left[\begin{array}{ccc}
	\widetilde{\vU}_{r} & \widetilde{\vU}'_{r} & \vU''_{n-2r}\end{array}\right]\underbrace{\left[
		\begin{array}{ccc}
		\vR_L   & \\
		& \vR_L \\
		&  & \vI_{n-2r}
		\end{array}
		\right]}_{\triangleq\vD_L} \in\mathbb{R}^{n\times n},
	\end{align*}
	are orthonormal bases for $\mathbb{R}^{n}$. Furthermore, we have
	\begin{equation*}
	\vB_L^T \widetilde{\vB}_L=
	\left[
	\begin{array}{ccc}
	\cos (\vGamma) & \sin (\vGamma)\\
	-\sin (\vGamma) & \cos (\vGamma)\\
	&  & \vI_{n-2r}
	\end{array}
	\right].
	\end{equation*}
	For $r$-dimensional subspaces ${\mathcal{V}}_r={\rm{span}}(\vX^T)$ and ${\widetilde{\mathcal{V}}}_r$, let $\vV_r$ and $\widetilde{\vV}_r \in \R^{n \times r}$ be orthonormal bases for ${\mathcal{V}}_r$ and ${\widetilde{\mathcal{V}}}_r$, respectively.
	Let the SVD of $\vV_r^T \widetilde{\vV}_r= \vL_R \cos (\vH) \vR_R^T$ with orthogonal matrices $\vL_R, \vR_R \in \R^{r \times r}$ and diagonal matrix $\vH \in\mathbb{R}^{r\times r}$, we use the same way to construct the orthonormal bases
	\begin{align*}
	\vB_R&=\left[\begin{array}{ccc}
	\vV_{r} & \vV'_{r} & \vV''_{n-2r}\end{array}\right]\underbrace{ \left[
		\begin{array}{ccc}
		\vL_R   & \\
		& \vL_R \\
		&  & \vI_{n-2r}
		\end{array}
		\right]}_{\triangleq\vC_R} \in\mathbb{R}^{n\times n} ,
	\nonumber\\
	\widetilde{\vB}_R &=\left[\begin{array}{ccc}
	\widetilde{\vV}_{r} & \widetilde{\vV}'_{r} & \vV''_{n-2r}\end{array}\right] \underbrace{ \left[
		\begin{array}{ccc}
		\vR_R   & \\
		& \vR_R \\
		&  & \vI_{n-2r}
		\end{array}
		\right]}_{\triangleq\vD_R} \in\mathbb{R}^{n\times n},
	\end{align*}
	such that
	\begin{equation*}
	\vB_R^T \widetilde{\vB}_R=
	\left[
	\begin{array}{ccc}
	\cos(\vH) & \sin(\vH)\\
	-\sin(\vH) & \cos(\vH)\\
	&  & \vI_{n-2r}
	\end{array}
	\right].
	\end{equation*}
	Similarly, the diagonal matrix $\vH= \diag\{\eta_1,\ldots,\eta_r\}\in\mathbb{R}^{r \times r}$ contains the principal angles between $\mathcal{V}_r$ and ${\widetilde{\mathcal{V}}}_r$ in a non-decreasing order.
\end{lemma}

Define the leverage scores of subspace
$\breve{\mathcal{U}}=\text{\rm{span}}([\vU_r,\widetilde{\vU}_r])$ and $\breve{\mathcal{V}}=\text{\rm{span}}([\vV_r,\widetilde{\vV}_r])$ as follows
\begin{align}
\breve{\mu}_{i}&\triangleq\mu_{i}(\breve{\mathcal{U}}),~i=1,2,\ldots,n,\label{def: mu_c}\\
\breve{\nu}_j&\triangleq\nu_j(\breve{\mathcal{V}}),~j=1,2,\ldots,n.\label{def: nu_c}
\end{align}

Under the assumptions of Lemma \ref{lem:canonical}, we can give the bound (or value) of $\alpha_1, \alpha_2, \xi_1$ and $\xi_2$ by using the following notations
\begin{align*}
\vA_{cc}&=\vL_L \cos(\vGamma) \vR_L^T \vR_R \cos(\vH) \vL_R^T,\\
\vA_{cs}&=\vL_L \cos(\vGamma) \vR_L^T \vR_R \sin(\vH) \vL_R^T,\\ \vA_{sc}&=\vL_L \sin(\vGamma) \vR_L^T \vR_R \cos(\vH) \vL_R^T,\\
\vA_{ss}&=\vL_L \sin(\vGamma) \vR_L^T \vR_R \sin(\vH) \vL_R^T.
\end{align*}

\begin{lemma} \label{lm: key_parameters}
	For $\vW_0= \vU_r\vV^T_r-\lambda \mathcal{P}_{\mathcal{T}}(\widetilde{\vU}_r\widetilde{\vV}^T_r)$ and $\mathcal{P}_{\mathcal{T}^{\bot}}(\widetilde{\vU}_r\widetilde{\vV}^T_r)$, it holds that
	\begin{align*}	
	&\norm{\vW_{0}}_F = \alpha_1,~\norm{ \lambda  \mathcal{P}_{\mathcal{T}^{\bot}}(\widetilde{\vU}_r\widetilde{\vV}^T_r)}
	=\alpha_2,\\
	&\norm{\vW_0}_{\mu(\infty,2)} \le \alpha_3 \beta,~	\norm{\vW_0}_{\mu(\infty)} \le \alpha_3 \beta,	
	\end{align*}
	where
	\begin{align*}
	\alpha_1^2&= 	\norm{\vI_r- \lambda \vA_{cc}}_F^2 + \norm{\lambda \vA_{cs}}_F^2 +\norm{\lambda \vA_{sc}}_F^2  \\
	\alpha_2 &=\norm{\lambda \vA_{ss}},~\alpha_3= \norm{\vI_r- \lambda \vA_{cc}} + \norm{\lambda \vA_{sc}} + \norm{\lambda \vA_{cs}},
	\end{align*}
	and
	$$
	\beta= 1 \vee \sqrt{2 \max_i\frac{\breve{\mu}_{i}}{\mu_{i}}} \vee \sqrt{2 \max_j\frac{\breve{\nu}_{i}}{\nu_{i}}}.
	$$
\end{lemma}
The proof of Lemma \ref{lm: key_parameters} is deferred to Appendix \ref{Proofoflemma8}.

\begin{remark} \label{Choice_of_Lambda_Genaral}
	In this case, the optimal choice of $\lambda$ is
	\begin{equation} \label{eq: Optimal_lambda}
	\lambda^\star=\frac{\tr(\vA_{cc})}{\norm{ \vA_{cc}}_F^2 + \norm{ \vA_{cs}}_F^2+ \norm{ \vA_{sc}}_F^2},
	\end{equation}
	which will achieve the minimum of $\alpha_1^2$
	$$
	\alpha_1^2 = r-\frac{\tr^2(\vA_{cc})}{\norm{ \vA_{cc}}_F^2 + \norm{ \vA_{cs}}_F^2+ \norm{ \vA_{sc}}_F^2}.
	$$
\end{remark}

A direct corollary of Lemma \ref{lm: key_parameters} for symmetric low-rank matrices is as follows.

\begin{lemma} For $\vW_0= \vU_r\vU^T_r-\lambda \mathcal{P}_{\mathcal{T}}(\widetilde{\vU}_r\widetilde{\vU}^T_r)$ and $\mathcal{P}_{\mathcal{T}^{\bot}}(\widetilde{\vU}_r\widetilde{\vU}^T_r)$, it holds that
	\begin{align*}	
	&\norm{\vW_{0}}_F = \alpha_1,~\norm{ \lambda  \mathcal{P}_{\mathcal{T}^{\bot}}(\widetilde{\vU}_r\widetilde{\vV}^T_r)}
	=\alpha_2,\\
	&\norm{\vW_0}_{\mu(\infty,2)} \le \alpha_3 \beta,~	\norm{\vW_0}_{\mu(\infty)} \le \alpha_3 \beta,	
	\end{align*}
	where
	\begin{align*}
	\alpha_1^2&=\lambda^2 \[r-\sum_{i=1}^{r} \sin^4\gamma_i\]-2 \lambda \sum_{i=1}^{r} \cos^2\gamma_i+r,\\
	\alpha_2&= \lambda \, \max_{i} \{\sin^2 \gamma_i\},\\
	\alpha_3&=\max_i \{1-\lambda \cos^2\gamma_i \}+2\lambda \, \max_i\{\cos\gamma_i \sin \gamma_i\},
	\end{align*}
	and
	$$
	\beta= 1 \vee \sqrt{2 \max_i\frac{\breve{\mu}_{i}}{\mu_{i}}}.
	$$	
\end{lemma}
\begin{IEEEproof} For symmetric matrix, we have $\vR_L^T \vR_R=\vI_r$, and then
	\begin{align*}
	\vA_{cc}&=\vL_L \cos(\vGamma) \cos(\vGamma) \vL_L^T, \vA_{cs}=\vL_L \cos(\vGamma) \sin(\vGamma) \vL_L^T,\\
	\vA_{sc}&=\vL_L \sin(\vGamma)  \cos(\vGamma) \vL_L^T, \vA_{ss}=\vL_L \sin(\vGamma)  \sin(\vGamma) \vL_L^T.
	\end{align*}
	By using the orthogonal invariance, we have
		\begin{align*}
		\alpha_1^2&=\norm{\vI_r- \lambda \cos(\vGamma) \cos(\vGamma)}_F^2+2\norm{\lambda \cos(\vGamma) \sin(\vGamma)}_F^2,\\
		\alpha_2& =\norm{\lambda\sin(\vGamma) \sin(\vGamma)},\\
		\alpha_3&= \norm{\vI_r- \lambda \cos(\vGamma) \cos(\vGamma)} + 2 \norm{\lambda \cos(\vGamma) \sin(\vGamma)}.	
		\end{align*}
	Incorporating the definition of $\vGamma$ completes the proof.
\end{IEEEproof}

Then by combining Theorem \ref{thm: main} and Lemma \ref{lm: key_parameters}, we achieve the following results.

\begin{theorem} \label{thm: main2-general}
	Let $\vX^\star \in \R^{n \times n}$ be a rank-$r$ matrix with thin SVD $\vX^\star=\vU_r \vSigma_r \vV^T_r$ for $\vU_r \in \R^{n \times r},\vV_r \in \R^{n \times r}$ and $\bm{\Sigma}_r \in \R^{r \times r}$. Let the column and row subspaces of $\vX^\star$ be ${\mathcal{U}}_r={\rm span}(\vU_r)$ and ${\mathcal{V}}_r={\rm span}(\vV_r)$, respectively. Assume that the $r$-dimensional prior subspace information ${\widetilde{\mathcal{U}}}_r$ about ${\mathcal{U}}_r$ and ${\widetilde{\mathcal{V}}}_r$  about ${\mathcal{V}}_r$ is known beforehand. Let $\vGamma \in \R^{r \times r}$ be diagonal whose entries are the principal angles between ${\mathcal{U}}_r$ and ${\widetilde{\mathcal{U}}}_r$ and $\vH \in \R^{r \times r}$ be diagonal whose entries are the principal angles between ${\mathcal{V}}_r$ and ${\widetilde{\mathcal{V}}}_r$. Let $\mu_i,\,\nu_{j},\,\breve{\mu}_{i}$ and $\breve{\nu}_{j}$ be the leverage scores defined as before.
	If
	\begin{multline*}
	1 \ge p_{ij}\gtrsim \max \left\{\log\(\frac{\alpha_1^2 n}{r\log n}\),1\right\} \cdot \frac{\left(\mu_{i}+\nu_{j}\right)r\log n}{n} \\
	\cdot \max\left\{\alpha_3^2\beta^2,1\right\}
	\end{multline*}
	for all $i,j=1,\ldots,n$, and
	$$ \alpha_2 < \frac{15}{16},
	$$
	then with high probability, we can achieve exact recovery of $\vX^\star$  by solving the program \eqref{OurMatrixCompletion2}, where
	\begin{align*}
	\alpha_1^2&= 	\norm{\vI_r- \lambda \vA_{cc}}_F^2 + \norm{\lambda \vA_{cs}}_F^2 +\norm{\lambda \vA_{sc}}_F^2  \\
	\alpha_2 &=\norm{\lambda \vA_{ss}},~\alpha_3= \norm{\vI_r- \lambda \vA_{cc}} + \norm{\lambda \vA_{sc}} + \norm{\lambda \vA_{cs}},
	\end{align*}
	and
	$$
	\beta= 1 \vee \sqrt{2 \max_i\frac{\breve{\mu}_{i}}{\mu_{i}}} \vee \sqrt{2 \max_j\frac{\breve{\nu}_{i}}{\nu_{i}}}.
	$$	
\end{theorem}

\section{Proof of Lemma \ref{lm: key_parameters}} \label{Proofoflemma8}
In this section, we will use principal angles between subspaces to bound $\norm{\vW_{0}}_{F},\norm{\vW_0}_{\mu(\infty)}$ and $\norm{\vW_0}_{\mu(\infty,2)}$ and $\mathcal{P}_{\mathcal{T}^{\bot}}(\widetilde{\vU}_r\widetilde{\vV}^T_r)$, where $\vW_0 = \vU_r\vV^T_r-\lambda \mathcal{P}_{\mathcal{T}}(\widetilde{\vU}_r\widetilde{\vV}^T_r)$. Before that, we give an alternative expression of $\vW_0$ first.  For convenience, we review the definition
\begin{align*}
	\vA_{cc}&=\vL_L \cos(\vGamma) \vR_L^T \vR_R \cos(\vH) \vL_R^T,\\
	\vA_{cs}&=\vL_L \cos(\vGamma) \vR_L^T \vR_R \sin(\vH) \vL_R^T,\\ \vA_{sc}&=\vL_L \sin(\vGamma) \vR_L^T \vR_R \cos(\vH) \vL_R^T,\\
	\vA_{ss}&=\vL_L \sin(\vGamma) \vR_L^T \vR_R \sin(\vH) \vL_R^T.
\end{align*}

We know $\vU_r^T\widetilde{\vU}_r=\vL_L \cos (\vGamma) \vR_L^T$ and $\vV_r^T \widetilde{\vV}_r =\vL_R \cos (\vH) \vR_R^T$. Besides, Lemma \ref{lem:canonical} immediately implies that
\begin{align*}
{\vU_r}&=\vB_{L}\left[\begin{array}{c}
\vL_L^T\\
\bm{0}_r\\
\bm{0}_{(n-2r)\times r}
\end{array}\right],\quad
\widetilde{\vU}_r=\vB_{L}\left[\begin{array}{c}
~~\cos(\vGamma) \vR_L^T\\
-\sin(\vGamma) \vR_L^T\\
~~\bm{0}_{(n-2r)\times r}
\end{array}\right],\\
{\vV_r}&=\vB_{R}\left[\begin{array}{c}
\vL_R^T\\
\bm{0}_r\\
\bm{0}_{(n-2r)\times r}
\end{array}\right],\quad
\widetilde{\vV}_r=\vB_{R}\left[\begin{array}{c}
~~\cos(\vH) \vR_R^T\\
-\sin(\vH)  \vR_R^T\\
~~\bm{0}_{(n-2r)\times r}
\end{array}\right].
\end{align*}

By incorporating the above expressions, we have
\begin{align*}
\mathcal{P}_{\mathcal{T}}(\widetilde{\vU}_r\widetilde{\vV}^T_r) &= \vU_r\vU_r^T \widetilde{\vU}_r\widetilde{\vV}^T_r + \widetilde{\vU}_r\widetilde{\vV}^T_r \vV_r \vV_r^T\\
&\qquad \qquad \qquad \qquad- \vU_r \vU_r^T \widetilde{\vU}_r\widetilde{\vV}^T_r \vV_r \vV_r^T\\
&=\quad \vB_{L} \vC_L^T\left[\begin{array}{ccc}
\vA_{cc} & -\vA_{cs} & \\
-\vA_{sc} & \bm{0}_r &\\
& &\bm{0}_{n-2r}
\end{array}\right]
\vC_R \vB^T_{R}.	
\end{align*}
So we have
\begin{align} \label{eq: W_0}
\vW_0&=\vU_r\vV^T_r-\lambda\mathcal{P}_{\mathcal{T}}(\widetilde{\vU}_r\widetilde{\vV}^T_r) \nonumber\\
&=\vB_{L} \vC_L^T \left[\begin{array}{ccc}
\vI_r- \lambda \vA_{cc} & \lambda \vA_{cs} & \\
\lambda \vA_{sc} & \bm{0}_r &\\
& &\bm{0}_{n-2r}
\end{array}\right]
\vC_R \vB^T_{R}.
\end{align}
1) \textbf{The new expression of $\norm{\vW_{0}}_{F}$}. Expressing $\vW_0$ by the principal angles \eqref{eq: W_0} yields
\begin{align*}
&\norm{\vW_{0}}_{F}	\\ &= \norm{\vU_r\vV^T_r-\lambda \mathcal{P}_{\mathcal{T}}(\widetilde{\vU}_r\widetilde{\vV}^T_r)}_F\\
&=\norm{\vB_{L} \vC_L^T \left[\begin{array}{ccc}
	\vI_r- \lambda \vA_{cc} & \lambda \vA_{cs} & \\
	\lambda \vA_{sc} & \bm{0}_r &\\
	& &\bm{0}_{n-2r}
	\end{array}\right]
	\vC_R \vB^T_{R}
}_F\\
&=\norm{\left[\begin{array}{ccc}
	\vI_r- \lambda \vA_{cc} & \lambda \vA_{cs} & \\
	\lambda \vA_{sc} & \bm{0}_r &\\
	& &\bm{0}_{n-2r}
	\end{array}\right]
}_F \\
&=\norm{\left[\begin{array}{cc}
	\vI_r- \lambda \vA_{cc} & \lambda \vA_{cs}  \\
	\lambda \vA_{sc} & \bm{0}_r
	\end{array}\right]}_F
\end{align*}
where the third equality holds due to the rotational invariance.
2) \textbf{The bound of $\norm{\vW_0}_{\mu(\infty)}$}. By using \eqref{eq: W_0}, the definition of $\vB_L$ and $\vB_R$ and the triangle inequality, we have
\begin{align*}
&\norm{\vW_0}_{\mu(\infty)} \\	
&\le \norm{\left(\frac{r \vM}{n}\right)^{-\frac{1}{2}}  \vU_r \cdot (\vI_r- \lambda \vA_{cc}) \cdot \vV_r^T \left(\frac{r \vN}{n}\right)^{-\frac{1}{2}}
}_\infty\\
&\quad + \norm{\left(\frac{r \vM}{n}\right)^{-\frac{1}{2}} \vU_r' \cdot \lambda \vA_{sc} \cdot \vV_r^T \left(\frac{r \vN}{n}\right)^{-\frac{1}{2}}}_\infty\\
&\quad +\norm{\left(\frac{r \vM}{n}\right)^{-\frac{1}{2}} \vU_r \cdot \lambda \vA_{cs} \cdot {\vV'_r}^T \left(\frac{r \vN}{n}\right)^{-\frac{1}{2}}}_\infty.
\end{align*}
By using
\begin{equation} \label{neq: bound_infty}
\norm{\vX\vY}_\infty \le \norm{\vX}_{(\infty,2)} \norm{\vY}_{(\infty,2)}
\end{equation}
and
\begin{equation} \label{neq: bound_infty_2}
\norm{\vX\vY}_{(\infty,2)} \le \norm{\vX}_{(\infty,2)} \norm{\vY},
\end{equation}
we have
\begin{align*}
&\norm{\vW_0}_{\mu(\infty)} \\
& \le \norm{\left(\frac{r \vM}{n}\right)^{-\frac{1}{2}} \vU_r}_{(\infty,2)} \norm{\vI_r- \lambda \vA_{cc}} \norm{\left(\frac{r \vN}{n}\right)^{-\frac{1}{2}} \vV_r}_{(\infty,2)}\\
& \quad + \norm{\left(\frac{r \vM}{n}\right)^{-\frac{1}{2}}  \vU_r'}_{(\infty,2)} \norm{\lambda \vA_{sc}} \norm{\left(\frac{r \vN}{n}\right)^{-\frac{1}{2}}\vV_r}_{(\infty,2)}\\
& \quad +\norm{\left(\frac{r \vM}{n}\right)^{-\frac{1}{2}}  \vU_r}_{(\infty,2)} \norm{\lambda \vA_{cs}} \norm{\left(\frac{r \vN}{n}\right)^{-\frac{1}{2}}\vV'_r}_{(\infty,2)}.
\end{align*}
Then we can obtain
\begin{align*}
	\norm{\vW_0}_{\mu(\infty)}  &\le \norm{\vI_r- \lambda \vA_{cc}}+ \norm{\lambda \vA_{sc}} \sqrt{2 \max_i\frac{\breve{\mu}_{i}}{\mu_{i}}}\\
	& \qquad \qquad \qquad \qquad
	+ \norm{\lambda \vA_{cs}} \sqrt{2 \max_j\frac{\breve{\nu}_{i}}{\nu_{i}}}\\
	& \le \left(\norm{\vI_r- \lambda \vA_{cc}}+ \norm{\lambda \vA_{sc}} + \norm{\lambda \vA_{cs}}\right) \cdot \\
	& \qquad \quad \( 1 \vee \sqrt{2 \max_i\frac{\breve{\mu}_{i}}{\mu_{i}}} \vee \sqrt{2 \max_j\frac{\breve{\nu}_{i}}{\nu_{i}}}\),
\end{align*}
 where the first inequality applies the following properties
\begin{align} \label{neq: groupbound}
\norm{\left(\frac{r \vM}{n}\right)^{-\frac{1}{2}} \vU_r}_{(\infty,2)}&=1,\,
\norm{\left(\frac{r \vN}{n}\right)^{-\frac{1}{2}} \vV_r}_{(\infty,2)}=1,\nonumber\\
\norm{\left(\frac{r \vM}{n}\right)^{-\frac{1}{2}} \vU_r'}_{(\infty,2)}&\le \sqrt{2 \max_i\frac{\breve{\mu}_{i}}{\mu_{i}}},\nonumber\\
\norm{\left(\frac{r \vN}{n}\right)^{-\frac{1}{2}}\vV_r'}_{(\infty,2)} &\le \sqrt{2 \max_j\frac{\breve{\nu}_{i}}{\nu_{i}}},
\end{align}
which are obtained by standard calculation \cite{eftekhari2018weighted}.

3) \textbf{The bound of $\norm{\vW_0}_{\mu(\infty,2)}$}. We recall the definition of $\norm{\vW_0}_{\mu(\infty,2)}$ as follows
\begin{multline*}
\norm{\vW_0}_{\mu(\infty,2)}\\= \norm{ \left(\frac{r \vM}{n}\right)^{-\frac{1}{2}} \vW_0} _{(\infty,2)}\vee \norm{\left(\frac{r \vN}{n}\right)^{-\frac{1}{2}} \vW_0^T} _{(\infty,2)}.	
\end{multline*}

Now we bound $\norm{ \left( r\vM/{n}\right)^{-{1}/{2}} \vW_0} _{(\infty,2)}$ first
\begin{align*}
&\norm{ \left(\frac{r \vM}{n}\right)^{-\frac{1}{2}} \vW_0} _{(\infty,2)}\\
&=\left \Vert \left(\frac{r \vM}{n}\right)^{-\frac{1}{2}} \vB_{L} \vC_L^T \left[\begin{array}{ccc}
\vI_r- \lambda \vA_{cc} & \lambda \vA_{cs} & \\
\lambda \vA_{sc} & \bm{0}_r &\\
& &\bm{0}_{n-2r}
\end{array}\right]
\right \Vert_{(\infty,2)},
\end{align*}
where the above equality uses the rotational invariance.
Using triangle inequality yields
\begin{align*}
&\norm{ \left(\frac{r \vM}{n}\right)^{-\frac{1}{2}} \vW_0} _{(\infty,2)}\\
&\le \quad\norm{ \left(\frac{r \vM}{n}\right)^{-\frac{1}{2}}
	\vU_r \cdot (\vI_r- \lambda \vA_{cc})}_{(\infty,2)} \\
&\quad + \norm{ \left(\frac{r \vM}{n}\right)^{-\frac{1}{2}}
	\vU_r' \cdot \lambda \vA_{sc}}_{(\infty,2)}\\
&\quad + \norm{ \left(\frac{r \vM}{n}\right)^{-\frac{1}{2}}
	\vU_r \cdot \lambda \vA_{cs}}_{(\infty,2)}.
\end{align*}
By using \eqref{neq: bound_infty_2} and \eqref{neq: groupbound}, we obtain
\begin{align*}
	&\norm{ \left(\frac{r \vM}{n}\right)^{-\frac{1}{2}} \vW_0} _{(\infty,2)}\\
	&\le \quad \norm{ \left(\frac{r \vM}{n}\right)^{-\frac{1}{2}}
		\vU_r}_{(\infty,2)} \norm{\vI_r- \lambda \vA_{cc}} \\
	&\quad+ \norm{ \left(\frac{r \vM}{n}\right)^{-\frac{1}{2}}
		\vU_r'}_{(\infty,2)} \norm{ \lambda \vA_{sc}}\\
	&\quad + \norm{ \left(\frac{r \vM}{n}\right)^{-\frac{1}{2}}
		\vU_r}_{(\infty,2)} \norm{ \lambda \vA_{cs}}\\
	&\le \quad \norm{\vI_r- \lambda \vA_{cc}} + \norm{ \lambda \vA_{sc}} \sqrt{2 \max_i\frac{\breve{\mu}_{i}}{\mu_{i}}} +  \norm{ \lambda \vA_{cs}}.
\end{align*}
Similarly, we have the other bound
\begin{multline*}
\norm{ \left(\frac{r \vN}{n}\right)^{-\frac{1}{2}} \vW^T_0} _{(\infty,2)} \le
\norm{\vI_r- \lambda \vA_{cc}}\\
+ \norm{ \lambda \vA_{sc}}
+  \norm{ \lambda \vA_{cs}}\sqrt{2 \max_i\frac{\breve{\nu}_{i}}{\nu_{i}}}.
\end{multline*}
Therefore, we obtain
\begin{align*}
\norm{\vW_0}_{\mu(\infty,2)} & \le \left(\norm{\vI_r- \lambda \vA_{cc}}+ \norm{\lambda \vA_{sc}} + \norm{\lambda \vA_{cs}}\right) \cdot \\
&\qquad  \qquad \( 1 \vee \sqrt{2 \max_i\frac{\breve{\mu}_{i}}{\mu_{i}}} \vee \sqrt{2 \max_j\frac{\breve{\nu}_{i}}{\nu_{i}}}\).
\end{align*}
4) \textbf{The new expression of $\norm{\lambda\mathcal{P}_{\mathcal{T}^{\bot}}(\vPhi)}$}. By applying rotational invariance, we obtain
\begin{align*}
&\norm{ \mathcal{P}_{\mathcal{T}^{\bot}}(\widetilde{\vU}_r\widetilde{\vV}^T_r)}
=\norm{\widetilde{\vU}_r\widetilde{\vV}^T_r-\mathcal{P}_{\mathcal{T}}(\widetilde{\vU}_r\widetilde{\vV}^T_r)}\\
&=\left\Vert\vB_{L} \vC_L^T \left[\begin{array}{ccc}
\vA_{cc} & -\vA_{cs} & \\
-\vA_{sc} & \vA_{ss} &\\
& &\bm{0}_{n-2r}
\end{array}\right] \vC_R\vB^T_{R} \right.
\\
&\quad\left.-
\vB_{L} \vC_L^T \left[\begin{array}{ccc}
\vA_{cc} & -\vA_{cs} & \\
-\vA_{sc} & \bm{0}_r &\\
& &\bm{0}_{n-2r}
\end{array}\right]
\vC_R\vB^T_{R}\right\Vert\\
&= \norm{
	\vB_{L} \vC_L^T \left[\begin{array}{ccc}
	\bm{0}_r & \bm{0}_r & \\
	\bm{0}_r & \vA_{ss} &\\
	& &\bm{0}_{n-2r}
	\end{array}\right]
	\vC_R\vB^T_{R}	} \\
&= \norm{\vA_{ss} }.
\end{align*}



\end{document}